\documentclass[10pt]{article}
\usepackage{amssymb}
\usepackage[dvips]{graphicx}
\usepackage{lscape}
\usepackage{multicol}

\begin{document}

\title{\bf ~\\
~\\~\\~\\~\\
ELECTRONIC STRUCTURE, BONDING AND OPTICAL SPECTRUM OF MgB$_2$}
\author{V.~P.~Antropov, K.~D.~Belashchenko, M.~van~Schilfgaarde$^{\dagger}$\\
and S.~N.~Rashkeev$^\ddagger$}
\date{}
\maketitle

\begin{center}
{\it Ames Laboratory, Iowa State University, Ames, IA 50011, USA\\[0pt]
$^\dagger$Sandia National Laboratories, Livermore, CA 94551, USA\\[0pt]
$^\ddagger$Vanderbilt University, Nashville, TN 37235, USA}
\end{center}

\section{Introduction}

The recent discovery of superconductivity (SC) in MgB$_{2}$~\cite{aki}
stimulated a significant interest in this system. One of the first questions
is whether MgB$_{2}$ represents a new class of superconductors, or whether
it may be related to other known classes in terms of its bonding and
electronic properties. Is it a unique system or just a representative of
some family of compounds with similar or even better properties?
The latter case seems unlikely after all studies~\cite{Slusky,Takenobu}
of possible dopings such as Mg$_{1-x}$Al$_{x}$B$_{2}$, MgB$_{2-x}$C$_{x}$,
Mg$_{1-x}$Li$_{x}$B$_{2}$ and Mg$_{1-x}$Mn$_{x}$B$_{2}$ showed loss of
SC, and many attempts to dope MgB$_{2}$ with other elements failed.

The crystal structure of MgB$_{2}$ may be regarded as that of completely
intercalated graphite~\cite{Bur} with carbon replaced by boron, its
neighbor in the periodic table. In addition, MgB$_{2}$ is formally
isoelectronic to graphite. Therefore, chemical bonding and electronic
properties of MgB$_{2}$ are expected to have strong similarity to those
of graphite and graphite intercalation compounds (GIC's), some of which
also exhibit SC.

The search for high-temperature SC in carbon phases started in 1965 when SC
with $T_{c}=0.55K$ in KC$_{8}$ was reported and subsequently explained in
terms of the interaction of $\pi$ bands with the longitudinal phonon modes
of the carbon layer~\cite{REVIEW}. Similar conclusions for the electronic
structure at the Fermi level were derived for LiC$_{6}$. The highest $T_{c}$
achieved for a GIC was 5~K~\cite{belash}. A parallel development was that of
SC in Bechgard salts, the organic charge-transfer systems. Within this
family of materials exhibiting carbon $\pi $ band conductivity, $%
T_{c}$ was raised to 12.5 K~\cite{12.5}. It should be noted that the
majority of these compounds are unstable at normal conditions and
high-pressure experiments are most common.

Later research shifted to the area of carbides of transition metals where
metallicity and conductivity are mostly due to transition metal atoms. Only
in 1991 was it found that alkali-doped fullerenes also exhibit SC with the
highest $T_{c}$ of 33~K~\cite{Tanigaki}. This group of 3D carbon-based
metals has a modified (compared to graphite) coupling of lattice vibrations
to Fermi electrons due to the curvature of the molecule, but the conducting
states still derive from the graphite $\pi$ band.

In this paper we will show that in spite of the structural similarity to
intercalated graphite and, so some extent, to doped fullerenes, MgB$_2$ has
a qualitatively different and rather uncommon structure of the conducting
states setting it aside from both these groups of superconductors. The
peculiar and (so far) unique feature of MgB$_2$ is the incomplete filling of
the two $\sigma$ bands corresponding to prominently covalent, $sp^2$%
-hybridized bonding within the graphite-like boron layer. The holes at the
top of these $\sigma$ bands manifest notably two-dimensional properties and
are localized within the boron sheets, in contrast with mostly
three-dimensional electrons and holes in the $\pi$ bands which are
delocalized over the whole crystal. These 2D covalent and 3D metallic-type
states contribute almost equally to the total density of states (DOS) at the
Fermi level, while the unfilled covalent bands experience strong interaction
with longitudinal vibrations of the boron layer.

Below we discuss the properties of MgB$_{2}$ following from
its theoretical treatment in the local density approximation (LDA). However,
an unusually strong non-linear electron-phonon coupling (EPC)~\cite{nonl}
together with the probable non-adiabaticity~\cite{Alexandrov}
raise a question whether the ground state and the excitation
spectrum may be adequately treated in LDA. In this situation it is
especially important to check the LDA predictions against the experimental
data related to the electronic structure of MgB$_{2}$ and doped alloys.
Therefore we also try to provide the basis for such comparison using both
our own results and those available in the literature.

This paper is organized as follows. The electronic structure of bulk and
surface of MgB$_{2}$ and its relation to GIC's is described in Section~2 along
with the effects of doping. Here we justify our choice of the
Mg$_{1-x}$Al$_{x}$B$_{2}$ system for detailed studies allowing one to
analyze trends associated with band filling and their relation to the
loss of SC. We believe that such analysis is more valuable
as opposed to the studies of a single system due to many uncertainties
(physical and numerical) of current band structure calculations and
experiments. The relevance of the theoretical DOS at the Fermi level
$N\equiv N(E_F)$ as well as the anisotropy of conducting states
are analyzed in Section~3 in connection with NMR data for
Mg$_{1-x}$Al$_{x}$B$_{2}$ alloy. Short
description of the EPC studies is presented in Section~4 along with the
comparison of theoretically predicted $N$ with that deduced from the
experimental data. Optical calculations for MgB$_2$ and AlB$_2$
single crystals are described in Section~5.

\section{Electronic structure and bonding}

MgB$_{2}$ occurs in the AlB$_{2}$ structure. Boron atoms reside in
graphite-like (honeycomb) layers stacked with no displacement~\cite{MGBSTR}
forming hexagonal prisms with the base translation almost equal to the
height, $a=3.085$ (3.009) \AA\ and $c/a=1.142$ (1.084) for MgB$_{2}$
(AlB$_{2}$). These prisms contain large, nearly spherical pores occupied
by Mg atoms.
As in graphite ($R_{\rm intra}$=1.42~\AA ), the intralayer B--B bonds are much
shorter than the interlayer distance, and hence the B--B bonding is strongly
anisotropic. However, the intralayer bonds are only twice as short as the
interlayer ones compared to the ratio of 2.4 in graphite, allowing for a
significant interlayer hopping. For comparison, the interatomic distance
between nearest neighbors is 1.55~\AA\ in diamond and 1.4--1.45~\AA\ in the C%
$_{60}$ molecule.

Early studies of the band structures of MgB$_{2}$ and
AlB$_{2}$ [12-15],
as expected, showed
their notable similarity to that of graphite (see e.g.~\cite{Freeman} and
references therein). Full-potential band structure calculations of AlB$_{2}$
are available in the literature~\cite{Freeman-review}. A recent
paper~\cite{Medv} reported the results of the studies
of MgB$_{2}$ and AlB$_{2}$, but the structure of conducting states was not
addressed.

Below we discuss the electronic structure of MgB$_{2}$ and some related
compounds obtained [18-20]
using the Stuttgart TB-LMTO (ASA) code,
full-potential LMTO (FLMTO)~\cite{nfpmethod} and full-potential LAPW (FLAPW)
methods. It was found that the
addition of gradient corrections to the local exchange-correlation potential
has only a tiny effect on the results. It also appears that a general
analysis of energy
bands for MgB$_{2}$ may be safely done in ASA, while a reliable treatment of
charge densities and anisotropy of transport properties requires more
accurate full-potential calculations. The band structure of AlB$_{2}$ is
much more sensitive to the choice of atomic spheres in ASA, and
full-potential treatment is imperative. Other methods were also used to
calculate the electronic structure of MgB$_{2}$ with very similar results.

\begin{figure}[tbp]
\begin{center}
\includegraphics*[scale=0.8]{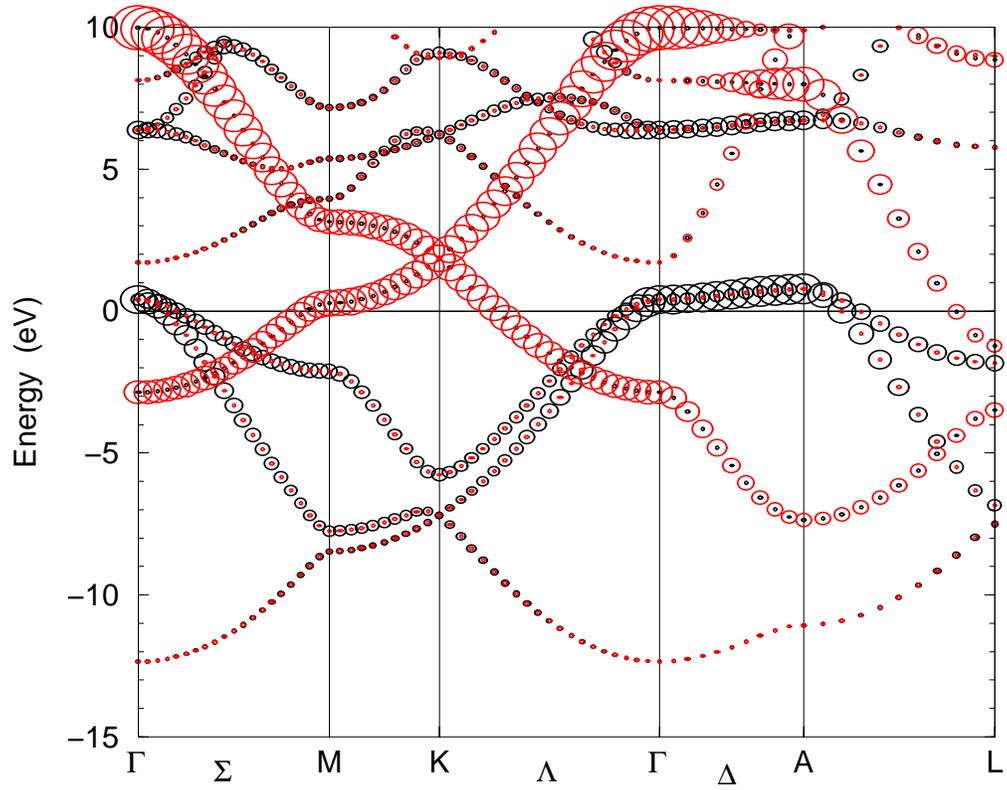}
\end{center}
\caption{Energy bands of MgB$_2$ from Ref.~\cite{us}. The radii of
black (red) circles are proportional to the boron $p_{xy}$ ($p_z$)
character.}
\label{mgb2bnds}
\end{figure}

\begin{figure}[tbp]
\begin{center}
\includegraphics*[scale=0.4]{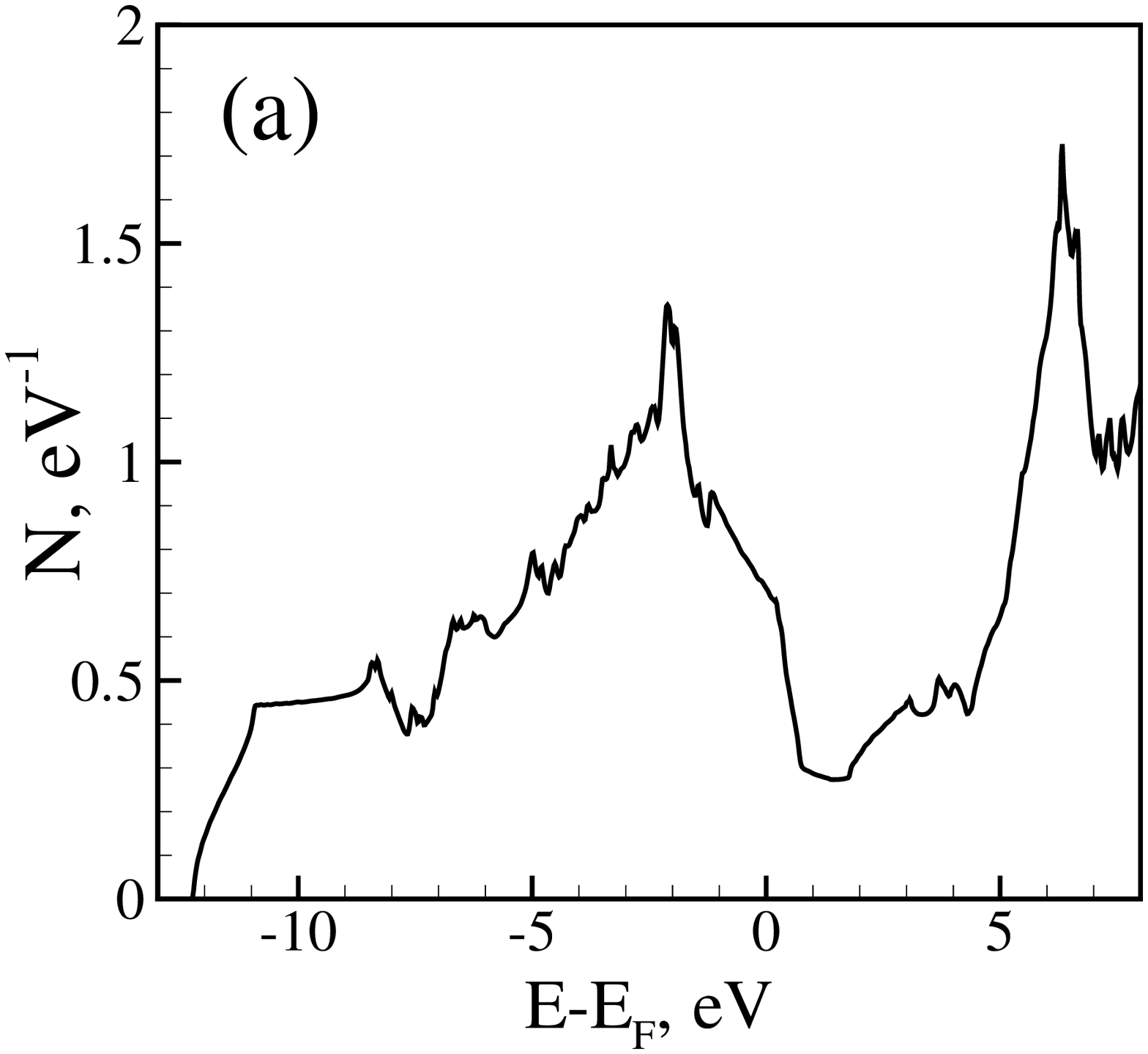}
\includegraphics*[scale=0.4]{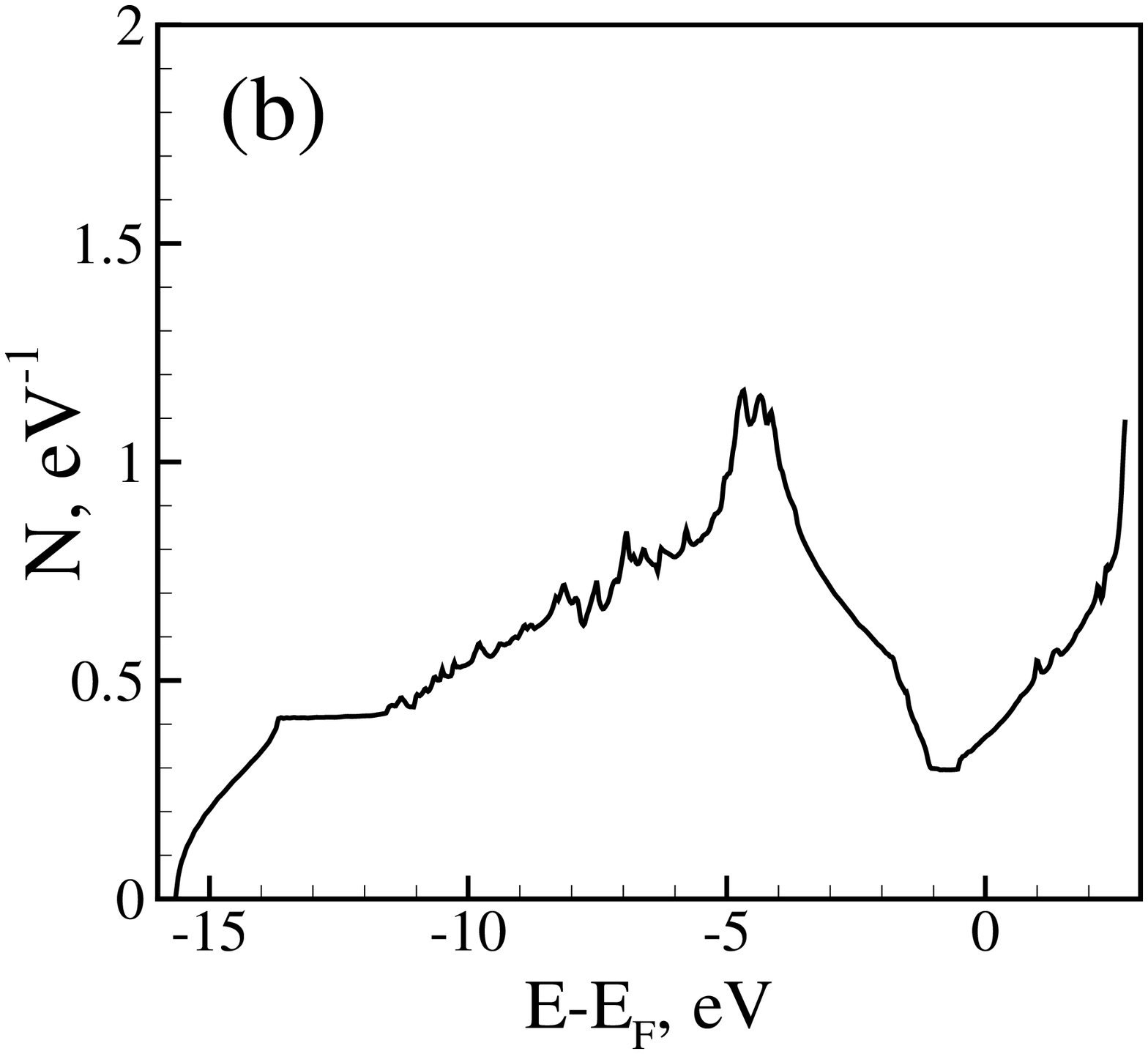}
\end{center}
\caption{Total density of states for (a) MgB$_2$ and (b) AlB$_2$.
Zero energy corresponds to the Fermi level.}
\label{DOS}
\end{figure}

\begin{landscape}
\begin{figure}[tbp]
\hskip10mm
\includegraphics*[scale=0.46]{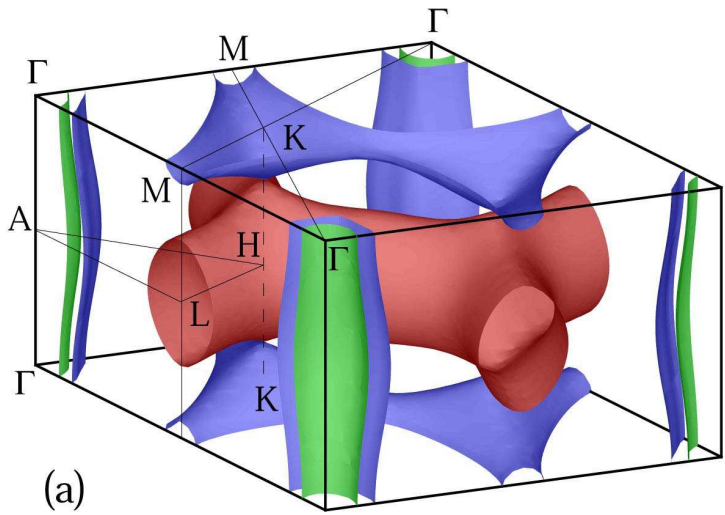}\hskip10mm
\includegraphics*[scale=0.51]{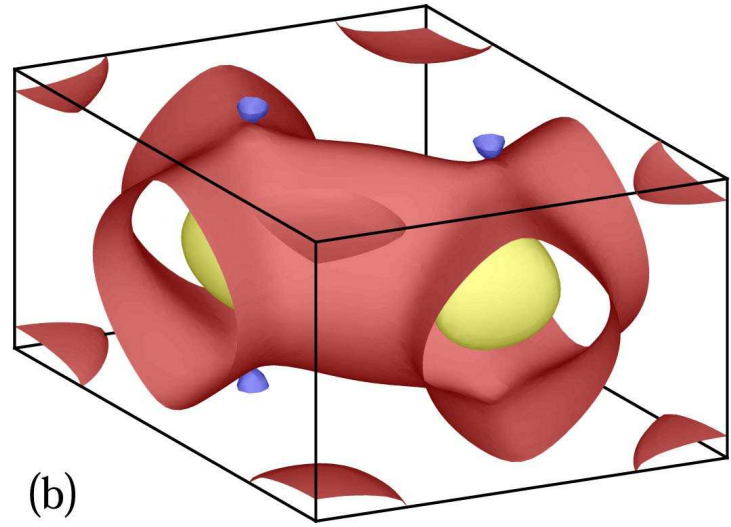}
\\
\vskip3mm
\hskip10mm
\includegraphics*{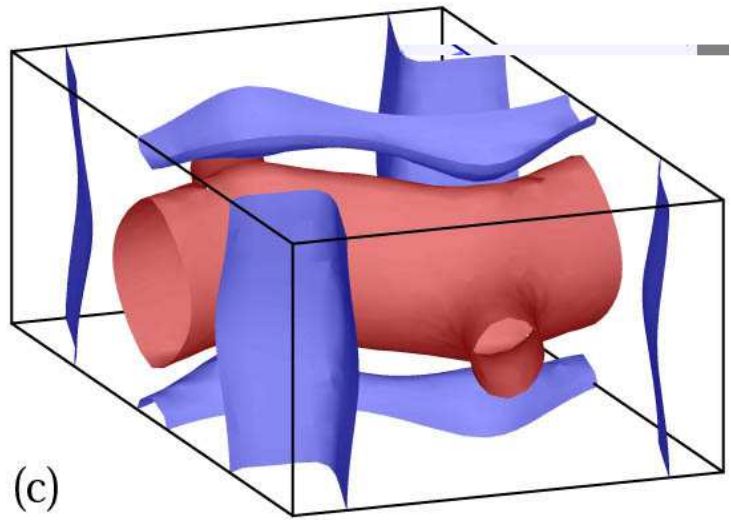}\hskip10mm
\includegraphics*{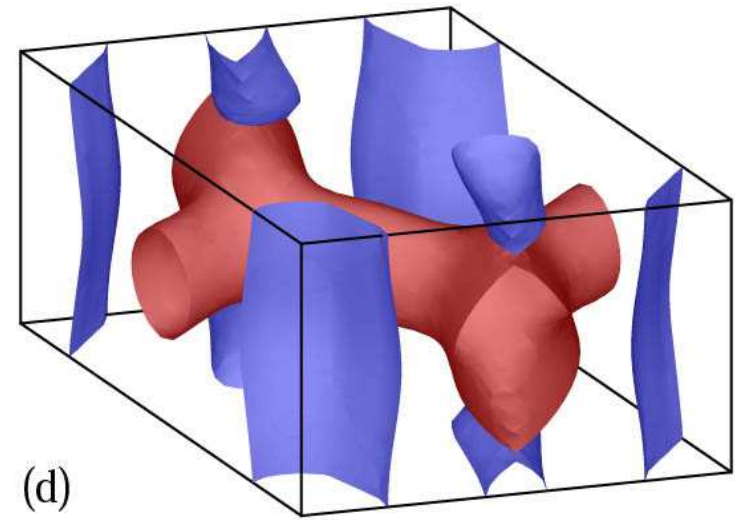}
\caption{Fermi surfaces of (a) MgB$_2$; (b) AlB$_2$; and
(c), (d) MgB$_2$ with $E_{2g}$ distortions of $\pm0.07a_0$
(see text).}
\label{FermiSurfaces}
\end{figure}
\end{landscape}

The energy bands, DOS and the Fermi surface of MgB$_{2}$ are shown in
Figs.~\ref{mgb2bnds},~\ref{DOS}a and~\ref{FermiSurfaces}a. As expected,
the bands are quite similar
to those of graphite with three bonding $\sigma $ bands corresponding to
in-plane $sp_{x}p_{y}$ ($sp^{2}$) hybridization in the boron layer and two
$\pi $ bands (bonding and antibonding) formed by aromatically hybridized
boron $p_{z}$ orbitals. Both $\sigma$ and $\pi$ bands have strong in-plane
dispersion due to the large overlap between all $p$ orbitals (both in-plane
and out-of-plane) for neighboring boron atoms. In particular, the total
width of the $\pi$ bands is 17~eV (which gives~\cite{Gunnarsson}
$\mu^{\ast}=0.14$, see below). The
interlayer overlaps are much smaller, especially for $p_{xy}$ orbitals, so
that the $k_{z}$ dispersion of $\sigma $ bands does not exceed 1~eV.
From the other hand, in contrast to graphite and GIC's,
two of the $\sigma $ bands are filled incompletely, as it was first noted in
Ref.~\cite{ArmstrongMgB2}. Together with weak $k_{z}$ dispersion this
results in the appearance of two nearly cylindrical sheets of the Fermi
surface (see Fig.~\ref{FermiSurfaces}a) around the $\Gamma $--A line.
As we will see below from
the analysis of the charge density distribution, these unfilled $\sigma $
bands with boron $p_{xy}$ character fully retain their covalent structure.
Conducting covalent bonds represent a peculiar feature of MgB$_{2}$ making it
an exotic compound probably existing on the brink of structural instability.
The blue (red) hole-type (electron-type) tubular network in
Fig.~\ref{FermiSurfaces}a corresponds to the bonding (antibonding) $\pi$ band.

It is seen in Fig.~\ref{FermiSurfaces}a that the $\pi$ bands form two
planar honeycomb tubular
networks: an electron-type sheet centered at $k_{z}=0$ (red) and another
similar, but more compact, hole-type sheet centered at $k_{z}=\pi /c$
(blue). These two sheets touch at some point on the K-H line. Note that the
hole-type sheet is close to the electronic topological transition (ETT) at
the M point corresponding to the breakdown of the tubular network into
separate shell-like pockets (at 0.25-0.30~eV above $E_{F}$). Although the
singularity in the electronic properties is weak, the proximity of this ETT
results in a strong coupling with the in-plane $E_{2g}$ phonons,
and vibrations of moderate amplitude are
able to break the links at the M point (see the discussion of
Figs.~\ref{FermiSurfaces}c and ~\ref{FermiSurfaces}d below). In addition,
this ETT takes place with electron doping (see below).

From Fig.~\ref{FermiSurfaces}a it is clear that the Fermi surface of MgB$_{2}$ has
characteristic features (the cylindrical sheets and the tubular links at M
and L points) that almost completely determine its topology. These features
are associated with relatively small electronic orbits that are especially
suitable for identification in de Haas-van Alphen experiments. Such
experiments on single crystals are highly desirable for the verification of
the band structure obtained in LDA.

\begin{figure}[tbp]
\begin{center}
\includegraphics*[scale=0.88]{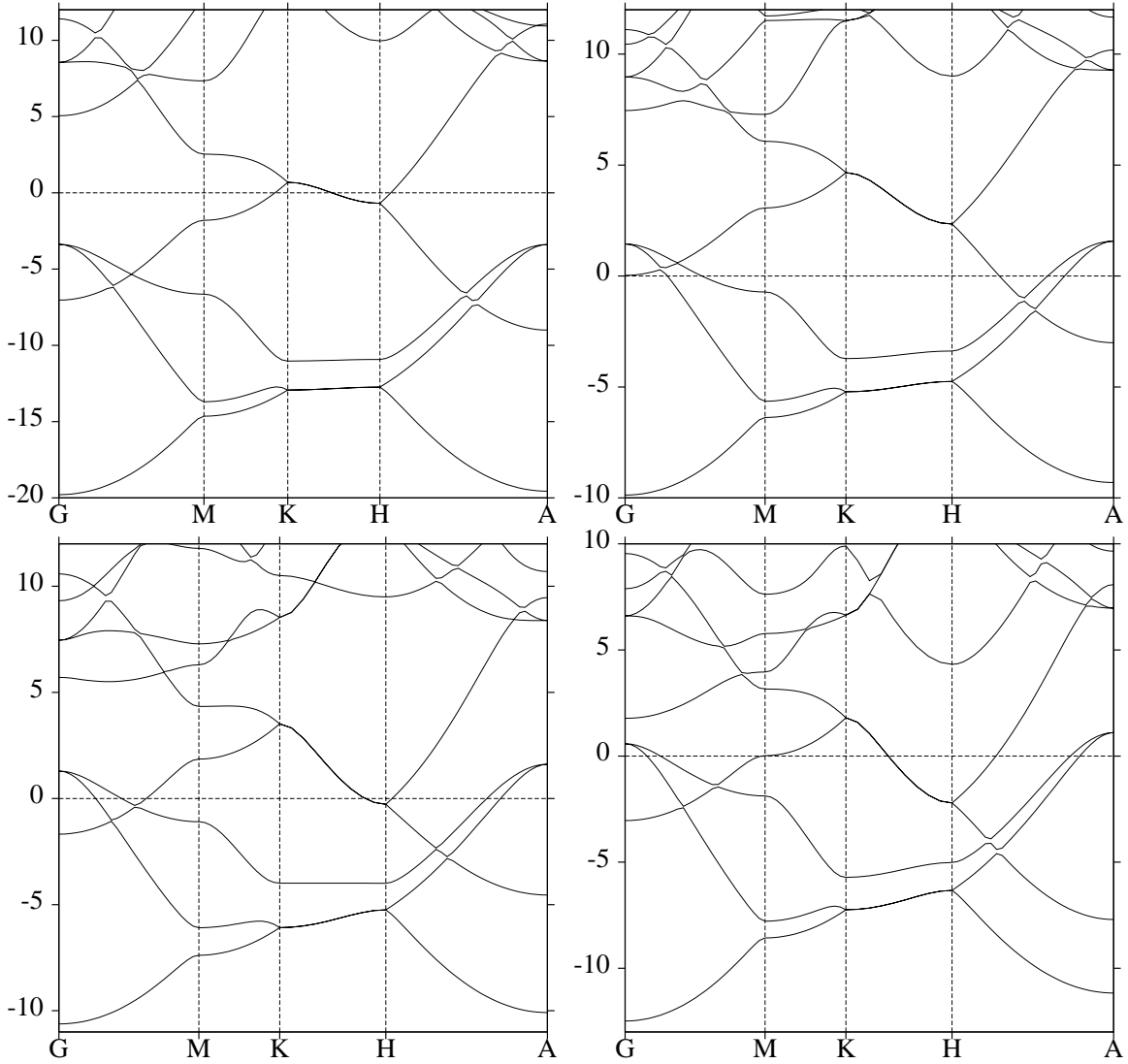}
\end{center}
\caption{Band structures of: (a) top left: primitive (AA stacking) graphite
(PG), $a=2.456$\AA, $c/a=1.363$; (b) top right: PG boron, $a=3.085$\AA,
$c/a=1.142$ (as in MgB$_2$); (c) bottom left: LiB$_2$ in MgB$_2$
structure, same $a$ and $c/a$; (d) bottom right: MgB$_{2}$, same $a$ and $%
c/a$. Energy is in eV relative to $E_{F}$. The order of occupied bands in
the $\Gamma $\ point is $\protect\sigma$ bonding with boron $s$ character,
$\pi$ bonding with boron $p_z$ character, and $\sigma$ bonding with boron
$p_{xy}$ character (double degenerate).}
\label{manybands}
\end{figure}

In order to examine the relation between the band structure of MgB$_{2}$\
and that of graphite in more detail we studied the following hypothetical
sequence of intermediate materials: carbon in the `primitive graphite' (PG)
lattice with no displacement between layers as in MgB$_{2}$,
using graphite lattice parameters; boron in the PG lattice with $a$\ as in
MgB$_{2}$\ and $c/a$ as in graphite; boron in the PG lattice with $a$\ and $%
c/a$\ as in MgB$_{2}$; LiB$_{2}$ in the same structure; MgB$_{2}$\ itself.
The results of some of these calculations are shown in Fig.~\ref{manybands}.

The band structure of PG carbon shown in Fig.~\ref{manybands}a is very similar to that of
graphite~\cite{Freeman} with the appropriate zone-folding for a smaller unit
cell. (This is quite natural because of the weak interlayer interaction.)
Boron in the same lattice dilated to match the MgB$_{2}$\ in-plane
lattice parameter (not shown) has nearly identical bands with the energies
scaled by the inverse square of the lattice parameter, in agreement with
common tight-binding considerations~\cite{Harrison}. Fig.~\ref{manybands}b shows the
natural enhancement of the out-of-plane dispersion of the $\pi $ bands when
the interlayer distance is reduced. Figs.~\ref{manybands}c and \ref{manybands}d demonstrate that
`intercalation' of boron by Li or Mg produces a significant distortion of
the band structure, so that the role of the intercalant is not simply one of
donating electrons to boron's bands (which would return the band structure
to that of PG carbon shown in Fig.~\ref{manybands}a). The main change upon intercalation
is the downward shift of the $\pi $ bands compared to $\sigma $ bands. For
Li this shift of $\thicksim $1.5 eV is almost uniform throughout the
Brillouin zone. Replacement of Li by Mg shifts the $\pi $ bands further, but
this shift is strongly asymmetric increasing from $\thicksim $0.6 eV at the $%
\Gamma $ point to $\thicksim $ 2.6 eV at the A point. In addition, the
out-of-plane dispersion of the $\sigma $ bands is also significantly
enhanced. In LiB$_{2}$ the filling of the bonding $p_{xy}$ bands is nearly
the same as in PG boron, while in MgB$_{2}$ the Fermi level shifts closer to
the top of these bands.

The lowering of the $\pi$ bands in MgB$_{2}$ compared to PG boron is due to
stronger interaction of boron $p_{z}$ orbitals with ionized magnesium
sublattice compared to $p_{xy}$ orbitals. This lowering is greater at the
AHL plane compared to the $\Gamma $KM plane, because the antisymmetric (with 
$k_{z}=\pi /c$) overlap of the boron's $p_{z}$ tails increases the
electronic density close to the magnesium plane where its attractive
potential is the strongest.

\begin{figure}[tbp]
\begin{center}
\includegraphics*[scale=0.45]{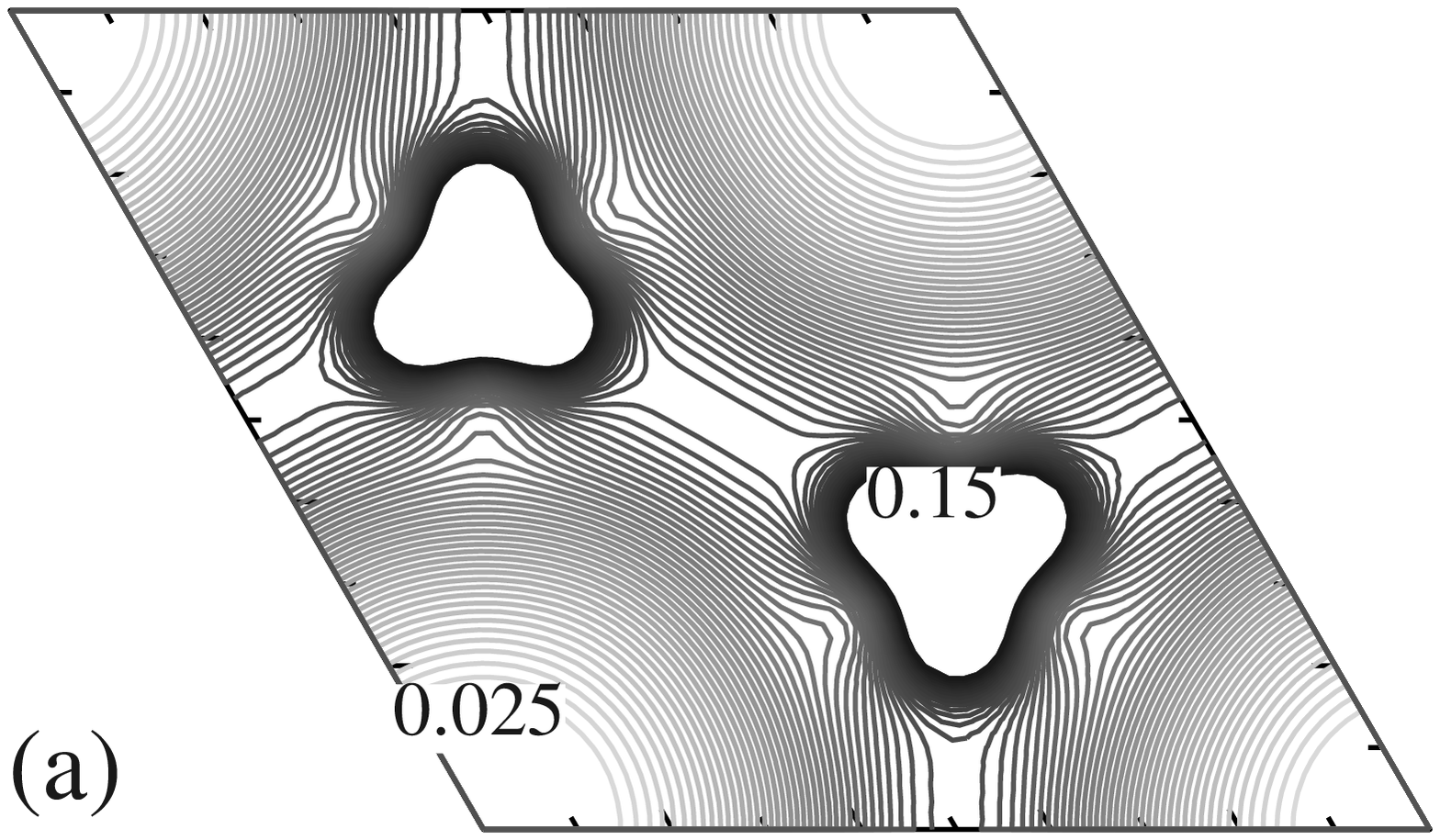}
\includegraphics*[scale=0.6]{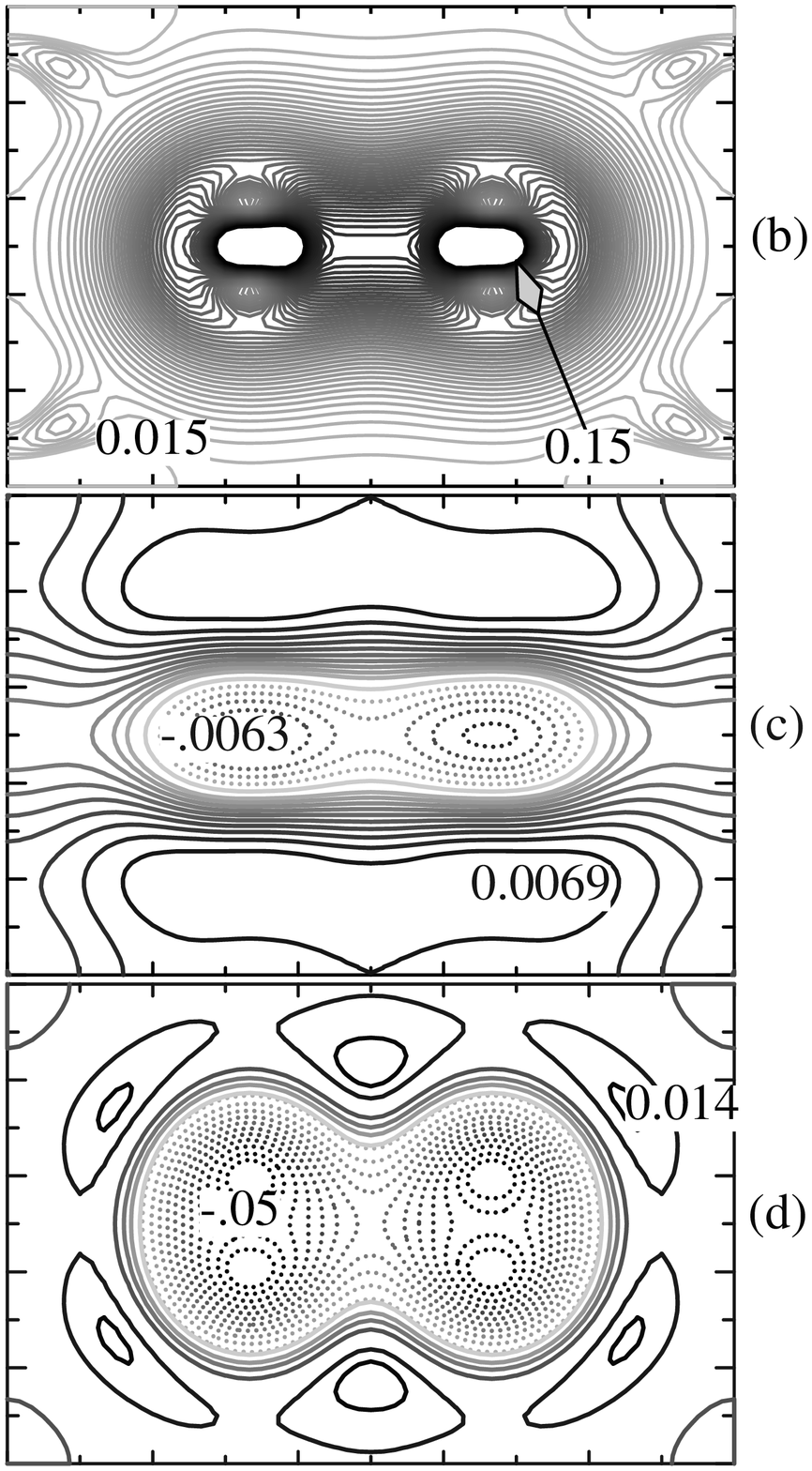}
\end{center}
\caption{Pseudocharge density contours obtained in FLMTO. The unit cell is
everywhere that of MgB$_{2}$. Darkness of lines increases with density. (a)
MgB$_{2}$\ in (0002) plane passing through B nuclei; (b) MgB$_{2}$\ in
(1000) plane passing through Mg nuclei at each corner of the figure. B
nuclei occupy positions (1/3,1/2) and (2/3,1/2) in the plane of the figure.
The integrated charge of the unit cell is 8. (c) (1000) plane, difference in
smoothed density, MgB$_{2}$\ minus NaB$_{2}$. The integrated charge of the
unit cell is 1. (d) (1000) plane, difference in smoothed density, MgB$_{2}$\
minus PG carbon. The integrated charge of the unit cell is 0. In (c) and
(d), dotted lines show negative values.}
\label{charge}
\end{figure}

The nature of bonding in MgB$_{2}$ may be understood from the charge density
(CD) plots shown in Fig.~\ref{charge}. As it is seen in Fig.~\ref{charge}a,
bonding in the boron layer is typically covalent. The CD of the boron atom
is strongly aspherical, and the directional bonds with high CD are clearly
seen (see also Ref.~~\cite{Medv}). The CD
distribution in the boron layer is very similar to that in the carbon layer
of graphite~\cite{Freeman}. This directional in-plane
bonding is also obvious from Fig.~\ref{charge}b showing the CD in the cross
section containing both Mg and B atoms. However, Fig.~\ref{charge}b also
shows that a large amount of valence charge does not participate in any
covalent bonding, but is rather distributed more or less homogeneously over
the whole crystal. Further, Fig.~\ref{charge}c shows the difference of the
CD of MgB$_{2}$\ and that of hypothetical NaB$_{2}$\ in exactly the same
lattice. Not only does it show that one extra valence electron is not
absorbed by boron atoms but is rather delocalized in the interstitials; it
also shows that some charge moves outward from boron atoms and covalent
in-plane B-B bonds. Fig.~\ref{charge}d shows the CD difference between the
isoelectronic compounds MgB$_{2}$\ and PG carbon (C$_{2}$). In MgB$_{2}$,
the electrons see approximately the same external potential as in C$_{2}$,
except that one proton is pulled from each C nucleus and put at the Mg site.
It is evident that the change C$_{2}${}$\rightarrow $MgB$_{2}$ weakens the
two-center $\sigma $\ bonds (the charge between the atoms is depleted) and
redistributes it into a delocalized, metallic density.

A numerical reconstruction of the electronic charge density from the
synchrotron radiation data for a powder MgB$_2$ sample~\cite{CDexp} supports
this general picture. The charge density found for 15~K is,
in fact, very similar to our Fig.~\ref{charge}b and shows all the important
features discussed above including the distinct covalent bonds within the
boron sheets, the strongly ionized Mg, and the delocalized charges in the
interstitials. Further, the Fourier maps obtained~\cite{MGBSTR} for the
single crystals also clearly show the covalent $sp^{2}$ hybrids in the boron
layer and no covalent bonding between B and Mg atoms.

\begin{figure}[tbp]
\begin{center}
\includegraphics*[scale=0.6]{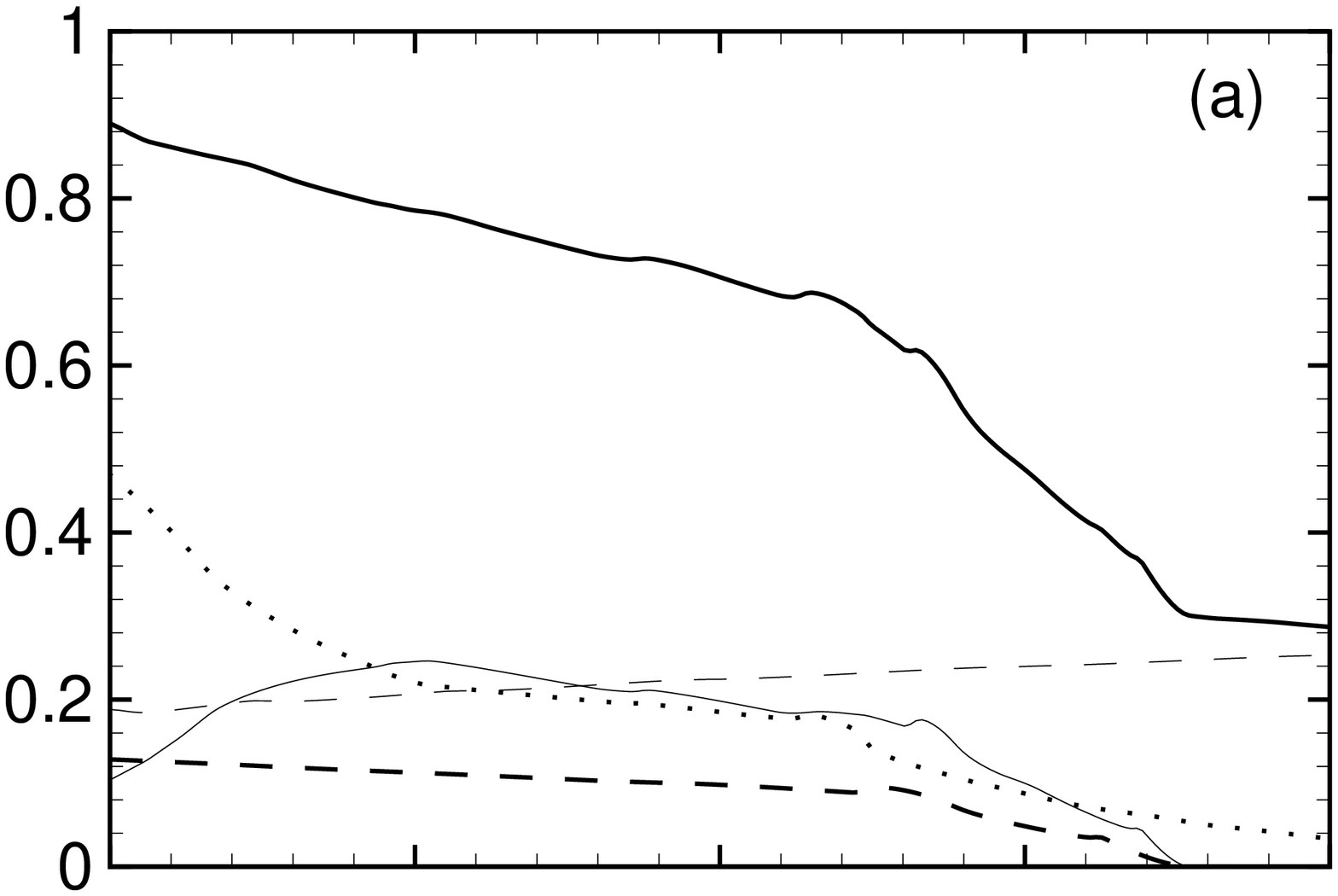}

\vskip1mm\hskip3mm
\includegraphics*[scale=0.6]{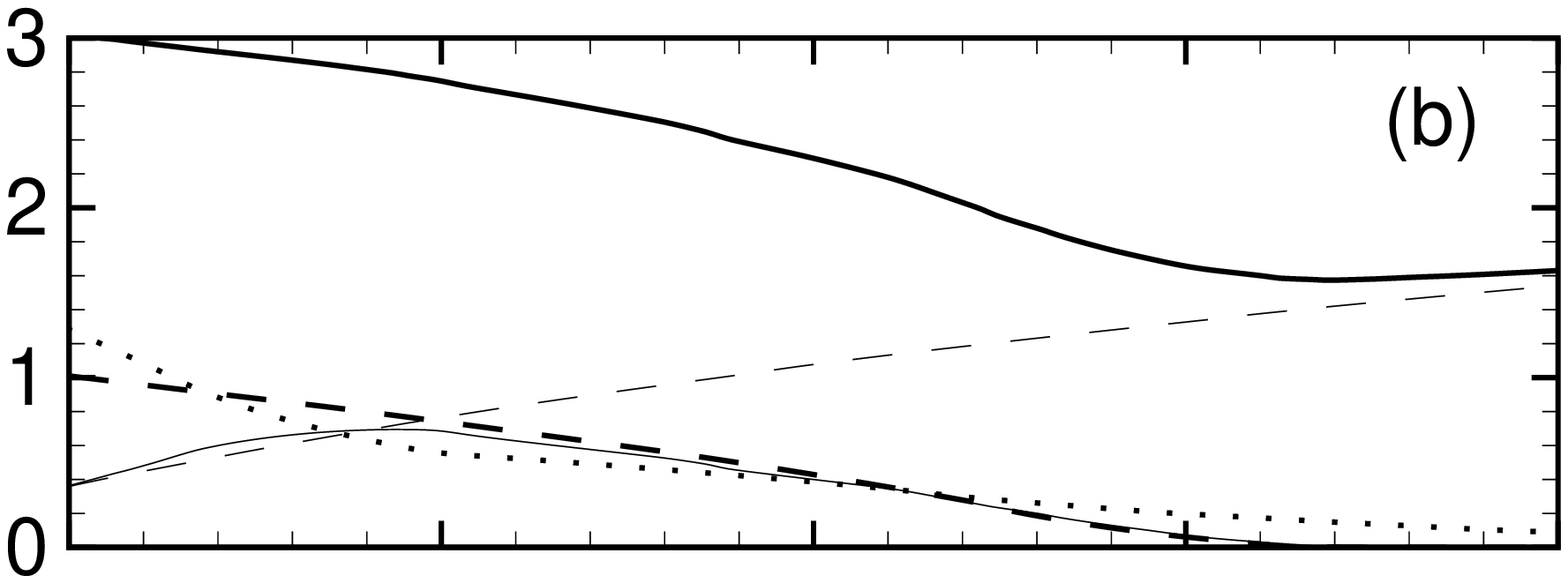}

\vskip2mm\hskip3mm
\includegraphics*[scale=0.6]{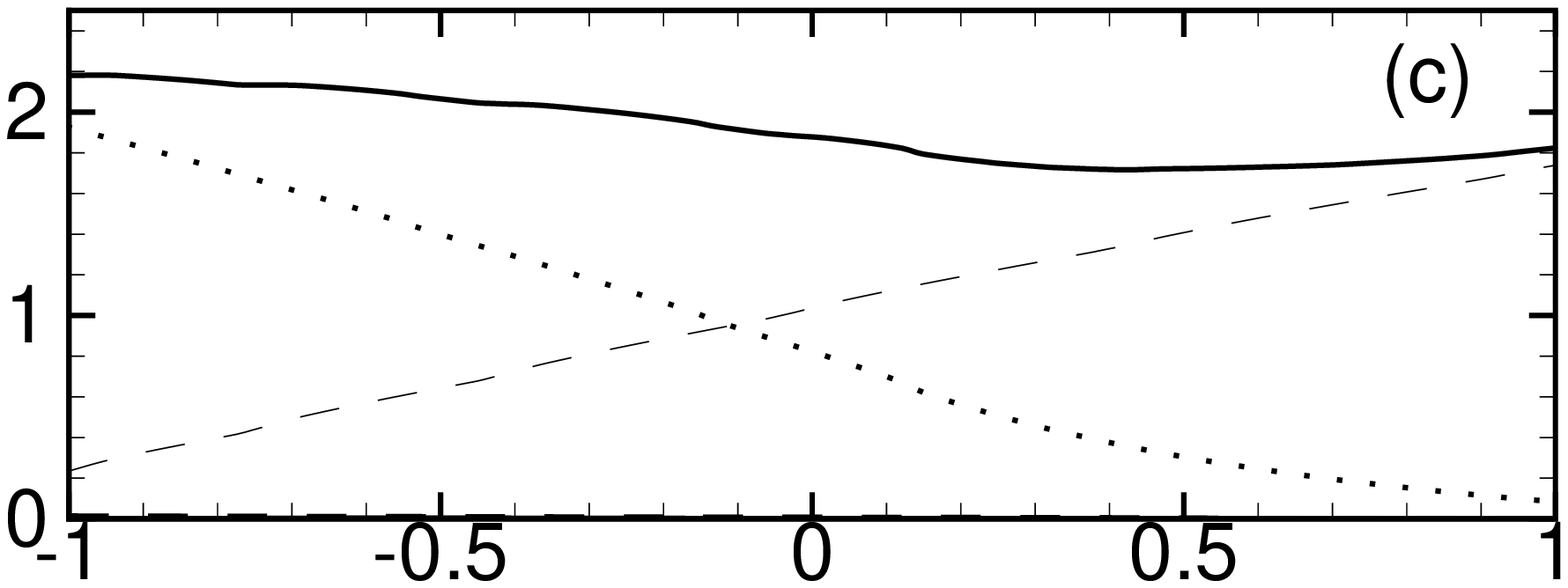}
\end{center}
\caption{Rigid band results for MgB$_{2}$: (a) Total DOS in eV$^{-1}$/cell;
(b) $\protect\sigma _{xx}$\ and (c) $\protect\sigma _{zz}$\ in
Ry$\cdot a_0^2$ (for definition of $\protect\sigma _{\protect\alpha 
\protect\beta }$\ see text), with contributions from different sheets of the
Fermi surface. Thick solid lines: total; thin dashed: 3D electronic sheet;
dotted: 3D hole sheet; thick dashed: internal cylinder; thin solid: external
cylinder. Energy is in eV relative to $E_{F}$.}
\label{decomposition}
\end{figure}

Because of the coexistence of two different types of conducting states, it
is desirable to find the contributions to the total DOS and transport
properties from separate sheets of the Fermi surface originating from 2D
covalent and 3D metallic-type bonding. Such decomposition is shown in
Fig.~\ref{decomposition} for the total DOS (see also Ref.~\cite{Liu})
and for the in-plane ($xx$) and
out-of-plane ($zz$) components of the tensor $\sigma_{\alpha\beta}=\int
v_{\alpha}v_{\beta}\delta(\varepsilon({\bf k})-E_F)d{\bf k}$, where $%
v_{\alpha}$ is the $\alpha$-component of the electronic velocity.
The 3D (metallic-type bonding) and cylindrical (covalent bonding)
parts of the Fermi surface contribute, respectively, about 55\% and 45\% to $%
N$. $N(E)$ for the hole-type zones rapidly decreases with
increasing $\varepsilon$ and already at $E-E_F\thickapprox0.8$~eV
the total DOS is almost completely determined by the 3D electron-type band.
The latter contribution is almost constant and probably not related to
the change of SC properties under pressure or with doping. The corresponding
contribution to conductivity exceeds all other contributions (more than 50\%
for $\sigma _{zz}$\ and $\sigma _{xx}$) and is virtually isotropic. Holes in
the two cylindrical sheets, as expected, have clearly anisotropic behavior
contributing nearly 30\% to $\sigma _{xx}$ and virtually nothing to
$\sigma_{zz}$. The 3D hole-type part of the Fermi surface is also notably
anisotropic with predominantly $z$-axis conductivity. The total $\sigma$ has
a rather small anisotropy at $E_F$ with $\sigma_{xx}/\sigma_{zz}\simeq1.22$.
The average projections of the Fermi velocities $\bar{v}_{\alpha }=\langle
v_{\alpha }^{2}\rangle ^{1/2}$ are: $\bar{v}_{x}=5.36\cdot 10^{7}$ cm/s, $%
\bar{v}_{z}=4.85\cdot 10^{7}$ cm/s (see also Refs.~\cite{us,Satta,An}).
The average Fermi velocities $\bar{v}%
_{\nu}=\langle v^{2}\rangle_{\nu }^{1/2}$ for each Fermi surface sheet $\nu$
are as follows: internal cylinder, 8.85$\cdot10^7$~cm/s; external cylinder,
6.02$\cdot10^7$~cm/s; 3D electronic sheet, 1.12$\cdot10^8$~cm/s; 3D hole
sheet, 8.74$\cdot10^7$~cm/s.

Thus, according to Ref.~\cite{us2}, the structure of MgB$_2$ is held
together by strongly {\it covalent}
bonding within boron layers and by delocalized, `{\it metallic}-type'
bonding between these sheets. A peculiar feature of this compound is that
electrons participating in both of these bond types provide comparative
contributions to $N$. This distinguishes MgB$_2$ from closely
related GIC's where covalent bonds in the carbon layers are always completely
filled, while the nearly cylindrical parts of the Fermi surface commonly
found in those compounds are formed by carbon-derived $\pi$ bands which are
also responsible for conductivity in pristine graphite~\cite{REVIEW}.

Thin films are of great technological interest; indeed probably the most
important application of MgB$_{2}$ at present is that for digital logic in
very high speed ($>$ 100 GHz) switching applications. Since the device speed
scales as the bandgap (and hence with $T_c$), MgB$_{2}$
shows great potential promise as the superconductor of choice. These devices
are grown as thin films; thus the role of the surface effects is of
interest.

\begin{figure}[tbp]
\begin{center}
\includegraphics*[scale=0.85]{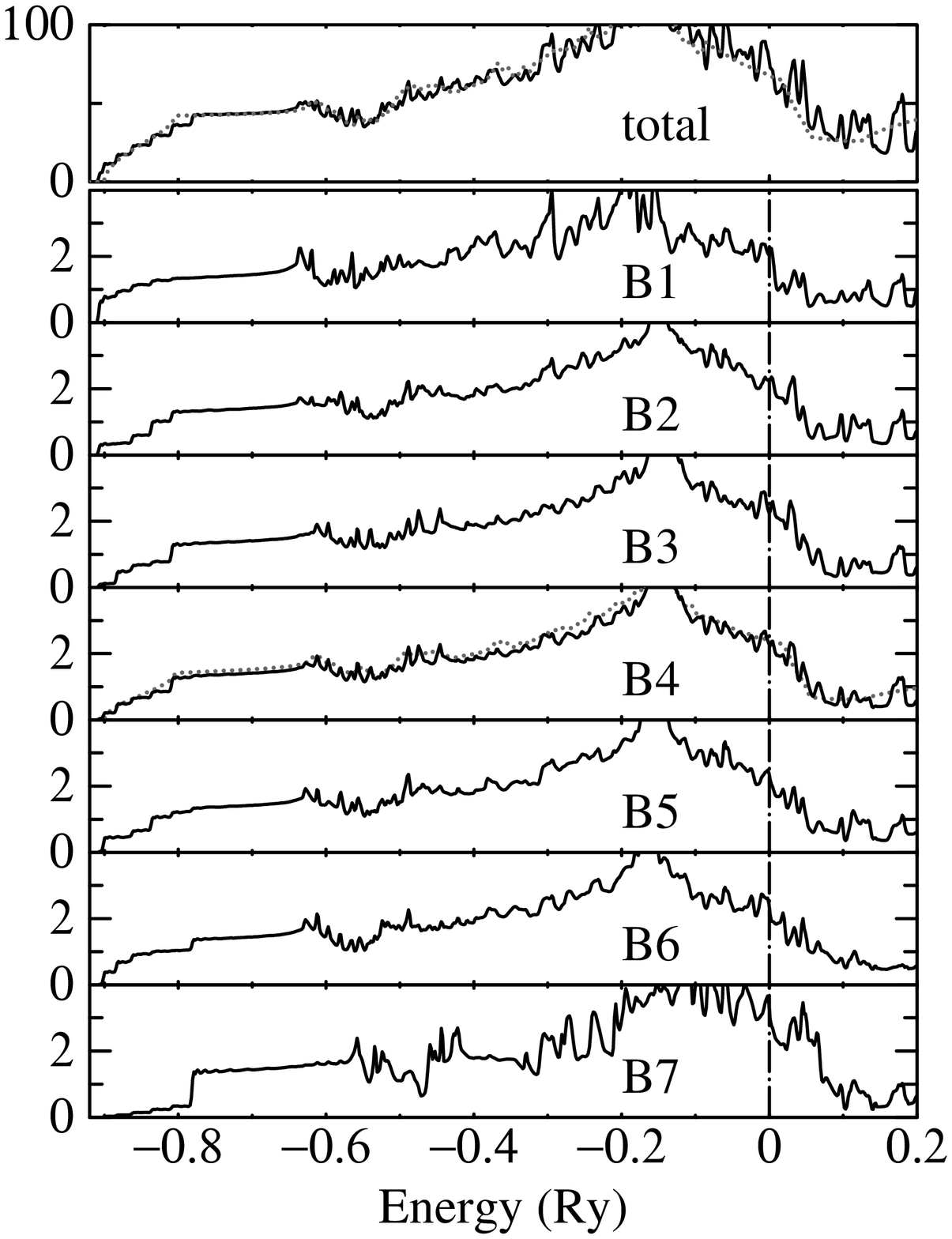}
\end{center}
\caption{Total DOS (top panel) and partial DOS in sites centered
at boron spheres, resolved by layer (panels 2-8) for 7-unit cell slab
of MgB2. The dotted lines show total (top panel) and partial
boron DOS (central panel) in bulk MgB$_2$. The first layer (B1)
corresponds to the Mg-terminated surface; the last (B7), to the
B-terminated surface (see text).}
\label{film}
\end{figure}

We considered a 7-unit cell (21 atom) slab of MgB$_{2}$, with 7
cells stacked along the $c$ axis, followed by a larger vacuum layer to
separate the two faces. A unit cell of MgB$_{2}$ consists of alternating Mg
and B planes along the $z$ axis. Thus this slab has two kinds of surfaces:
a Mg-terminated surface at $z=0$ and a B-terminated surface at $z=7c$ for an
ideal, unrelaxed geometry. The slab was fully relaxed; however, it retains
the hexagonal symmetry of the bulk lattice, so the only allowed
relaxations are in the $z$ coordinate (no symmetry-lowering reconstruction
was considered). A substantial relaxation was found at both surfaces, each
contracting inward so as to reduce the spacing between layers. The
relaxation was most pronounced at the Mg-terminated surface (the surface
Mg-B spacing contracted by 12\%). Fig.~\ref{film} shows total DOS (top panel) and
partial DOS in sites centered at boron spheres, resolved by layer (panels
2-8). The DOS shown was computed for 8000 points in the full Brillouin zone,
using the electronic structure method of Ref.~\cite{nfpmethod}. The dotted
lines correspond to that of bulk MgB$_{2}$: 7 times the total DOS of one
unit cell of MgB$_{2}$ (top panel) and B partial DOS (central panel). The
first layer corresponds to the Mg-terminated surface; the last to the
B-terminated surface. Several points are evident. (1) The B partial DOS of
the central layer is quite similar to that of the bulk DOS, showing that
surface perturbations heal in just a few monolayers, as expected for a
metal. (2) There is a shift in the B partial DOS towards higher energies at
the B-terminated surface (B7), and to lower energies at the Mg
terminated surface (B1). To a large extent this effect is electrostatic: the
electrostatic potential at the MT boundary for B1 is about 32~mRy higher
than for the central layer B4, while the corresponding potential for B7 is about
9~mRy lower. (3) There is an enhancement of DOS near the Fermi level at the
B-terminated surface. This is in part due to the electrostatic shift, but also
there is an additional structure in the local DOS corresponding to the formation
of the surface band which localizes approximately at B6 and B7. (4) There is
a slight increase in the total DOS relative to the bulk (top panel), which
originates in the surface states from the B-terminated surface.

Now let us discuss the relation of MgB$_2$ to other compounds. The closest
existing material is the isostructural AlB$_2$ corresponding to
the addition of one electron to MgB$_2$. Al is the only neighbor of Mg in
the periodic table that may be used as a dopant producing an isostructural
solid solution with a reasonably wide single-phase region. As it is always
more useful to study a sequence of similar compounds or doping trends in
alloys instead of analyzing a single system, below we discuss the trends
in the band structure and related properties of the Mg$_{1-x}$Al$_x$B$_2$
alloy.

\begin{figure}[tbp]
\begin{center}
\includegraphics*[scale=0.67]{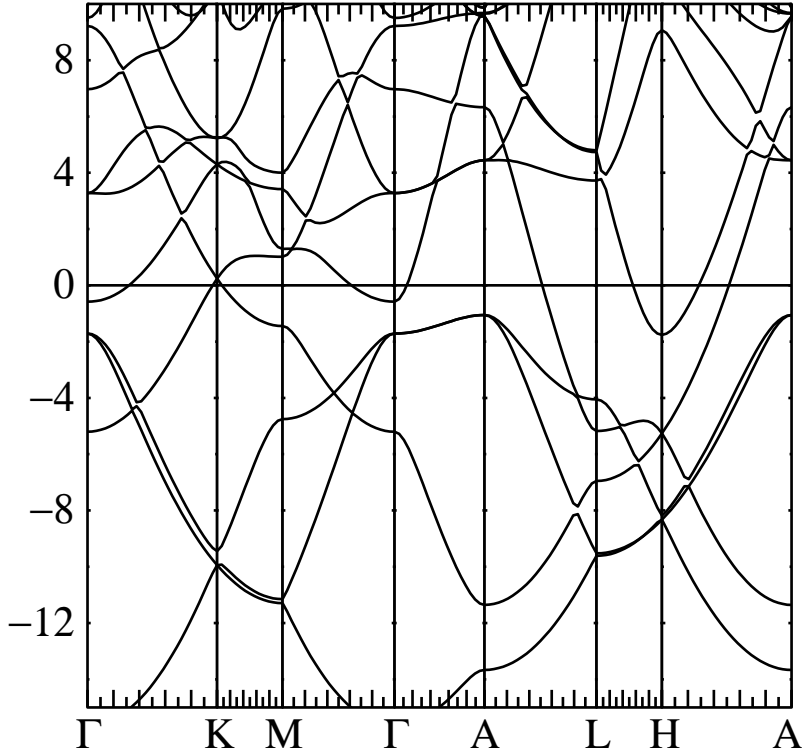}
\end{center}
\caption{Energy bands for AlB$_2$.}
\label{AlB2-BS}
\end{figure}

Our FLMTO band structure and DOS of AlB$_2$ shown in Figs.~\ref{AlB2-BS}
and~\ref{DOS}b are in excellent agreement with earlier
results~\cite{Freeman-review,Medv}.

The Fermi surface for AlB$_2$ is shown in Fig.~\ref{FermiSurfaces}b.
The red electron-type network (blue pocket at $K$)
corresponds to the antibonding (bonding) $\pi$ band, while red and
yellow pockets at $\Gamma$ and $H$ correspond to the `interstitial'
electron-type band. In contrast to MgB$_2$, the
$\sigma$ bands in AlB$_2$ are completely filled, and the Fermi surface has no
cylindrical sheets. This difference stems from an extra valence electron,
but the rigid band picture does not fully describe the effect of
replacement of Mg by Al~\cite{Satta}. The $\pi$ bands are pushed further
down in the AHL plane (compared to the $\sigma$ bands), together with the
interstitial band~\cite{Satta} which is unfilled in MgB$_2$ ($\sim$2~eV at
the $\Gamma$ point). In AlB$_2$ the latter interstitial band falls below the
Fermi level and forms electronic pockets at $\Gamma$ and H points. The
hole-type tubular sheet of the Fermi surface found in MgB$_2$ shrinks to
very small hole pockets at the K point in AlB$_2$.

It has been shown~\cite{Slusky} that Mg$_{1-x}$Al$_x$B$_2$ in the $0\leq
x\leq0.4$ region forms single-phase solutions at $0\leq x\lesssim0.1$ and at 
$0.25\lesssim x$ with the same AlB$_2$ structure, while at $0.1\lesssim
x\lesssim0.25$ the system appears to decompose in two isostructural
AlB$_2$-type phases with different $c/a$ ratios. The Mg-rich phase remains
superconducting up to $x\simeq0.1$ where $T_c$ is only 2~K lower than in
pure MgB$_2$, while the transition broadens in the two-phase region and
there is no supeconductivity in the Al-rich phase with $x\gtrsim0.25$ (as
well as in pure AlB$_2$~\cite{ALB2}).

Unfortunately, very little is known about the subtle structural transition
in the Mg$_{1-x}$Al$_x$B$_2$ alloy. However, as this transition is
accompanied by the loss of SC, more detailed experimental
characterization of the structural transition as well as of the
dependence of the electronic structure on doping are highly desirable.
In particular, our FLMTO calculations show that in the rigid band model at
approximately $x=0.17$ the band structure undergoes an above-mentioned ETT
at the M point where the hole-type tubular sheet of the Fermi surface splits
into compact pockets. In principle, this ETT may be the source of the
structural transition, but verification of this connection needs further
studies.

Hole doping by replacing Mg by Na or Li seems logical, but the
experiments on Mg$_{1-x}$Li$_{x}$B$_{2}$ (as well as on
MgB$_{2-x}$C$_{x}$ and Mg$_{1-x}$Mn$_{x}$B$_{2}$) revealed the
loss of SC~\cite{Takenobu}.
Weak SC was observed under
pressure in NaC$_{2}$~\cite{NAC2} which is very unstable. However, it may be
possible to form an alloy (Mg,Na)B$_{2}$\ with modest amounts of Na. It is
evident from Fig.~\ref{decomposition} that the hole-type parts of the Fermi
surface change dramatically with band filling.

The band structure of BeB$_2$ in AlB$_2$ lattice calculated with experimental
average or optimized lattice parameters is quite similar to that of
MgB$_2$~\cite{Freeman-review,Satta}. The differences include a notable
reduction of the lattice parameters (especially of the $c/a$ ratio), wider
valence bands, larger dispersion of the $\sigma$ bands, and a somewhat
different shape of the $\pi$ bands~\cite{Satta}. Due to this latter
difference, the hole-type and electron-type sheets of the Fermi surface
corresponding to bonding and antibonding $\pi$ bands become almost identical
in BeB$_2$. The ETT at the M point (tubular link breaking) shifts
approximately 2~eV higher compared to MgB$_2$ and becomes irrelevant
for the EPC.

The hypothetical ZnB$_{2}$ in MgB$_{2}$ structure is also very similar
to MgB$_{2}$ in terms of the electronic properties. Our calculations for
this compound with lattice parameters of MgB$_{2}$ produced a very similar
band structure with a nearly identical Fermi surface.
Another recently discovered
isoelectronic system LiBC~\cite{LIBC} contains both B and C. According to
our results this system is a perfect insulator, and any substitution of C by
B will lead to metallic behavior. Experimentally small conductivity was
observed in LiBC~\cite{LIBC}. Electronic structure of this system is also
very similar to that of BN~\cite{BN}.

\section{Nuclear spin-lattice relaxation}

For SC, one of the important parameters is $N$ in the normal state.
Several groups have performed
LDA calculations of this quantity and the results
range within 0.68--0.74 states/(eV$\cdot $f.u.) [18-20,~26].
Experimentally $N$\ can be determined in many different ways but the nuclear
spin-lattice relaxation (NSLR) rate $T^{-1}_{1}$ measurements represent an
excellent opportunity to check experimentally not only the total $N$ and its
partial components but also their anisotropy, i.e. the distribution between
the in-plane and out-of-plane $p$\ orbitals. For instance, in the studies of 
$N$ in alkali-doped C$_{60}$ the theoretical analysis of NMR data provided a
qualitatively new interpretation of NMR experiments.

For NSLR on
$^{11}$B ($\mu =2.689\mu_{N}$) in MgB$_{2}$ the
experimental papers~[31-33]
reported $TT_{1}=$180, 155 and 165 K$\cdot$sec.
The relaxation rates were interpreted in terms of
dipolar and orbital contributions due to the low Korringa ratio and
the known dominance of $p$ states at the Fermi level~[18-20,~26].
Below we show the results obtained in a more careful theoretical analysis
for Mg$_{1-x}$Al$_x$B$_{2}$~\cite{USNMR}. As we will see, the studies of
NSLR in this system may provide valuable information about the genesis of
anisotropy of electronic states at the Fermi level.

We used the following expressions for a monocrystaline material with
the hexagonal symmetry~\cite{OBATA}:
\[
(T_{1}^{-1})_{F}^{s}=(4\pi kT/\hbar )(h\gamma _{N}H_{F})^{2}N_{s}^{2} 
\]
\[
(T_{1}^{-1})_{\rm orb}^{p}=(4\pi kT/\hbar )(h\gamma _{N}H_{\rm orb}^{p})^{2}\left[
2N_{E^{\prime }}N_{A_{1}^{\prime \prime }}-N_{E^{\prime }}(N_{E^{\prime
}}-N_{A_{1}^{\prime \prime }})\sin ^{2}\theta \right] 
\]
for Fermi-contact and orbital contributions, respectively. Here $%
H_{\rm orb}^{p}=2\mu _{B}\langle r^{-3}\rangle _{p}$ and $H_{F}=(8\pi /3)\mu
_{B}[\varphi _{s}(0,E_{F})]^{2}/4\pi $ are the corresponding hyperfine
fields, $\langle r^{-3}\rangle _{p}$ is the expectation value of $r^{-3}$
over the $p$ state $\varphi _{p}(r,E_{F})$, and $N_{\Gamma }$ is the partial
DOS at the Fermi level for the representation $\Gamma $ (for more details
see Refs.~\cite{OBATA,USC60}). To obtain these values we used the
LMTO-ASA method. To check the sensitivity of
our results to the parameters of calculations we used different
exchange-correlation potentials and inputs with different radii of the
B sphere (both with and without empty spheres). 
The relevant partial contributions to $N$ in
MgB$_{2}$ and AlB$_{2}$ are listed in Table~\ref{tabledos}. One can see that
in MgB$_{2}$
all $p$ orbitals on the B site have a sizeable DOS, while in AlB$_{2}$
only $p_{z}$ orbital has a large DOS with $N_{px}\approx 0.1N_{pz}$. The $s$
component in AlB$_{2}$ becomes relatively more important compared to
MgB$_{2}$ resulting in the dominance of the Fermi-contact mechanism
of NSLR, as we will show
below. In both materials the contribution of $d$ states to NSLR is very
small. As for the Mg site, the $s$ component of $N$ is the most important,
and we expect that NSLR for the Mg nucleus is dominated by the Fermi-contact
mechanism. However, below we will focus on the $^{11}$B NSLR, because of the
boron role in SC.

\begin{table}[tbp]
\caption{Partial DOS for $s$ and $p$ orbitals at B site, 10$^{-3}$ (eV$\cdot 
$spin$\cdot $atom)$^{-1}$}
\begin{center}
\begin{tabular}{|l|c|c|c|}
\hline
\ & $s$ & $p_z$ & $p_x$ \\
\hline
MgB$_{2}$ & 3.4 & 50 & 36 \\ 
\hline
AlB$_{2}$ & 3.3 & 19 & 1.9\\
\hline
\end{tabular}
\end{center}
\label{tabledos}
\end{table}

To calculate $T_{1}$ according to a general prescription one has
to estimate the values of $\langle r^{-3}\rangle_l$ for different $l$ and
the electronic density at the nucleus $\varphi_s^2(0)/4\pi$. We found that
the convergence of the total NSLR rate with respect to the boron sphere radius
in this $sp$ system is worse compared to $d$ metals (from $r_B=2.3a_0$ to
$r_B=2.4a_0$ the total $T^{-1}_1$ decreases by $\sim$15\%). The
uncertainty is mostly related to the value of $\langle r^{-3}\rangle_{p}N_p$.

In our calculations we used the largest $r_B$ that were possible without
a significant distortion of the band structure, 2.4$a_0$ for MgB$_{2}$
and 2.1$a_0$ for AlB$_{2}$. For these radii we have
$\langle(a_0/r)^3\rangle_{p}=1.11$ in MgB$_2$ and 1.37 in AlB$_2$.
For comparison, the atomic value~\cite{Fraga} for
$\langle (a_0/r)^3\rangle_{p}$ in B is 0.78. The electronic densities on
the nucleus $a_0^3\varphi_{s}^{2}(0)/4\pi$ for MgB$_2$ and AlB$_2$ were,
respectively, 2.68 and 3.02. 

The contributions to the $^{11}$B relaxation rate for the polycrystalline
sample calculated using the general formulas~\cite{OBATA} 
are given in Table~\ref{tableNMR}.
The in-plane and out-of-plane $p$ orbitals in MgB$_{2}$ have similar
DOS, and hence the relative magnitude of orbital and dipolar
contributions to NSLR is close to the 3/10 rule for $p$ states in a
cubic crystal described by Obata~\cite{OBATA}.
The Fermi-contact
contribution is also important and amounts to $\sim$40\% of the orbital term.
The contributions from the $d$ partial waves to the dipole and orbital
relaxation rates are small (at the order of 1\%) due to the low diagonal
and off-diagonal densities of states $(N_d/N_p)^2\sim0.02$ and 
$(N_{pd}/N_p)^2\sim0.05$. The quadrupole contribution to NSLR is negligible
due to the rather small $^{11}$B quadrupole moment.

\begin{table}[tbp]
\caption{Contributions to $(TT_{1})^{-1}$ [$10^{-4}$ (K$\cdot$sec)$^{-1}$]}
\begin{center}
\begin{tabular}{|l|c|c|c|c|c|c|}
\hline
\  & Contact & Orbital & Dipole & Total & Enhanced & Experiment
\\ \hline
MgB$_{2}$ & 12 & 30 & 9 & 51 & 81 & 56~\cite{EXP1}, 64~\cite{EXP2},
61~\cite{EXP3}\\ 
AlB$_{2}$ & 21 & 1 & 1 & 23 & 26 & ---\\
\hline
\end{tabular}
\end{center}
\label{tableNMR}
\end{table}

The values of $T^{-1}_{1}$ obtained in such manner correspond to the
theoretical `bare' partial boron $N$ whereas the actual $T^{-1}_1$
contains different terms enhanced according to the corresponding susceptibilities.
To estimate the possible range of enhancement, we calculated the total
enhancement coming from the spin susceptibility only (which is the case
for AlB$_2$).
We estimated the effective Stoner exchange parameter
$I\equiv\Delta E/m=1.7$~eV from the splitting of the bands at the
$\Gamma$ point in the external magnetic field. The corresponding Stoner
enhancement of the uniform spin susceptibility in the 3D case may be
written~\cite{ENHANS} as
$S=3/\left[(1-NI)\left(3-2IN\right)\right]\approx(1-IN)^{-\alpha}$
with $\alpha\approx1.62$, while in the 2D case the enhancement is described by
the same formula with $\alpha=2$. In our case due to the mixed
2D and 3D character of the bands it is not clear what value of $\alpha$
should be used, but the difference in the result for
$\alpha=1.62$ and 2.0 is less than 10\%. We used $\alpha=1.9$ resulting
in the enhancement of $T_1^{-1}$ by approximately 60\% (Table~\ref{tableNMR}).
The obtained `bare' and enhanced values provide a range of
possible NSLR rates for MgB$_2$ which should be compared with the
experimental rates~[31-33]
of (56--64)$\cdot10^{-4}$
(K$\cdot$sec)$^{-1}$. 
The fact that such simple estimate may give a faster relaxation
compared to experiments suggests a possible importance of unique
effects resulting in the lowering of the effective boron $N$.

The roles of the three NSLR mechanisms are very different in AlB$_2$ where
no experimental data are
available. According to our theoretical estimation, due to the sharp decrease of
the boron $p$ component of $N$ compared to MgB$_2$ and its very strong anisotropy
(see Table~\ref{tabledos}), the orbital and dipolar contributions to NSLR become very small,
and the NSLR in AlB$_2$ is completely dominated by the Fermi-contact
mechanism. This conclusion may be verified experimentally by studying the
Korringa ratio. The resulting Stoner-enhanced NSLR rate in AlB$_2$ is more than
three times smaller than in MgB$_2$ (see Table~\ref{tableNMR}).

\begin{figure}[tbp]
\begin{center}
\includegraphics*[scale=0.5]{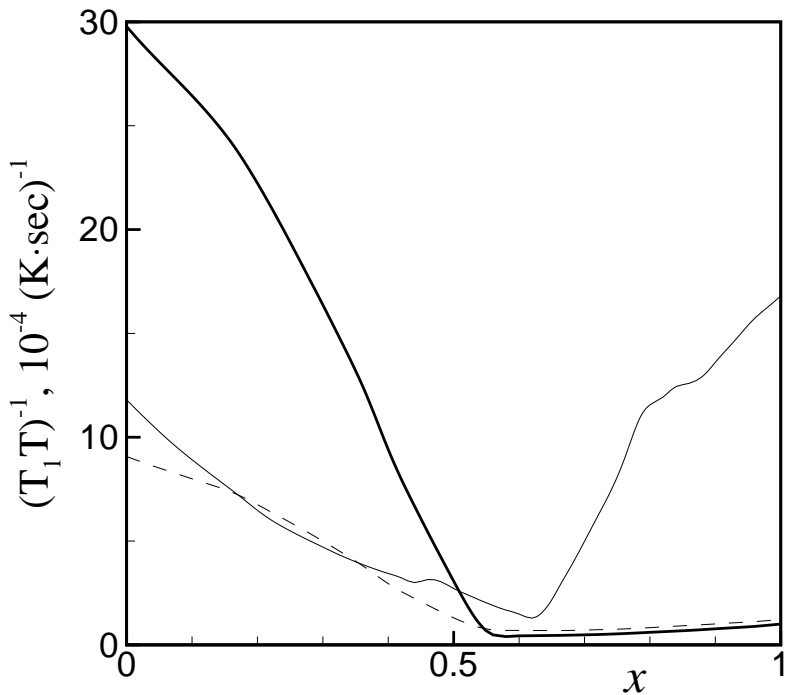}
\end{center}
\caption{Different contributions to the total $^{11}$B NSLR
in Mg$_{1-x}$Al$_{x}$B$_{2}$. Thick solid line, orbital;
dashed, dipolar; thin solid, Fermi-contact.}
\label{FigNSLR}
\end{figure}

We also calculated the NSLR rate in Mg$_{1-x}$Al$_x$B$_2$ in the rigid band
approximation with MgB$_2$ bands. From Fig.~\ref{FigNSLR} one can see how the roles of
different mechanisms of NSLR change with doping. The sharp decrease of $N$
in the 2D sheets of the Fermi surface with doping~\cite{us2} leads to the
corresponding lowering of all contributions to NSLR, and at the point of the
complete filling of these 2D bands we expect a very large $T_1$.
Experimental verification of this sharp dependence of $T_1$ on doping
in this alloy may be a crucial test of our understanding of the electronic
structure of this system and is highly desirable. Together with the nuclear
quadrupole resonanse data (which is related to the anisotropy of total
charges on different $p$ orbitals) such measurements should help to build
a general picture of the anisotropy of $p$ orbitals.

The above calculations have been done for a polycrystalline sample.
Because single crystals are becoming available, we include our estimations
of the anisotropy in the angular dependence of NSLR rate~\cite{OBATA}
$A+B\sin^2\theta$. For MgB$_2$ we obtained $B/A\approx-0.06$, so that
the NSLR is nearly isotropic. The NSLR rate in AlB$_2$ is isotropic
because it is determined by isotropic Fermi-contact mechanism.

Summarizing the results of this Section, the orbital NSLR mechanism
dominates over the spin-dipolar and Fermi-contact ones in MgB$_2$, because
the boron $p$ orbitals at the Fermi level are distributed nearly isotropically 
and have a large DOS.
Strong anisotropy and low density of $p$ states at the Fermi level
in AlB$_2$ results in the dominance of the Fermi-contact mechanism.
With the values of 0.7 (0.5) st./eV for the bare total
(boron partial) DOS at the Fermi level for MgB$_2$ and 1.7~eV for the
effective Stoner exchange parameter, reasonable agreement is obtained with
the experimental NSLR rates. In addition, the NSLR rate strongly depends
on the concentration in the Mg$_{1-x}$Al$_x$B$_2$ system, and the
experimental studies of this dependence are highly desirable
for the understanding of the anisotropy of boron $p$ states.
We note a relatively weak radial convergence
of the NSLR rate in this system compared to the well studied $d$ metals,
as well as a common uncertainty related to the estimate of the
enhancement factor. However, these uncertainties do not affect our
qualitative conclusions.

Similar results for NSLR in MgB$_{2}$\ were obtained in Ref.~\cite{Mazin},
where $T_{1}$\ on $^{25}$Mg and the Knight shifts were also computed.
We believe that the discrepancy in the Fermi-contact term is mainly due
to the larger boron sphere radius used in Ref.~\cite{USNMR}.

\section{Electron-phonon coupling}

Already in the first publication~\cite{us} following the discovery of
SC in MgB$_2$ the strength of the EPC was estimated and a
qualitative suggestion that MgB$_2$ is a standard BCS superconductor was
made. Measurements of the B-isotope effect~\cite{Budko} on T$_{c}$,
tunneling~\cite{Sharoni}, transport~\cite{tra}, thermodynamic
properties~\cite{otherheat}, and the phonon density of
states~\cite{PHODOS} confirm
that MgB$_{2}$ is most likely an electron-phonon mediated $s$-wave
superconductor with intermediate or strong coupling.

\begin{figure}[tbp]
\begin{center}
\includegraphics*[scale=0.5]{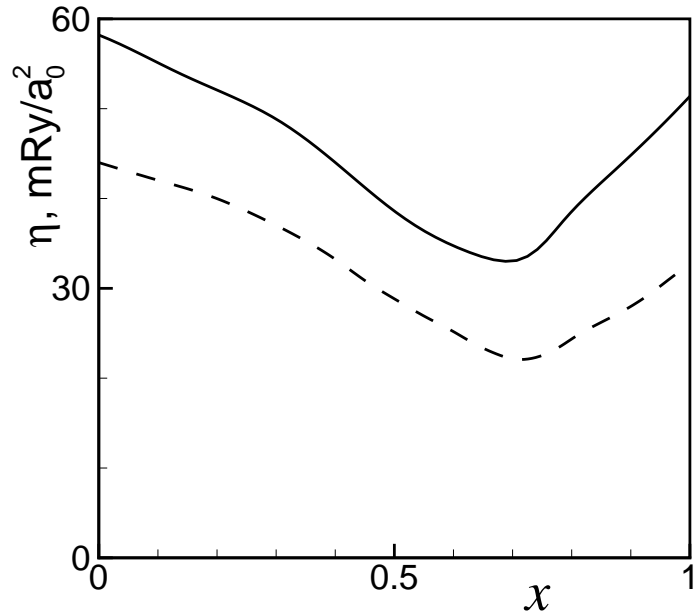}
\end{center}
\caption{Rigid band results for the Hopfield parameter $\eta$
in Mg$_{1-x}$Al$_{x}$B$_{2}$ obtained in
non-spherical rigid-ASA approximation.
Rigid band shift starting from MgB$_2$ (solid line), and
from AlB$_2$ (dashed line).}
\label{eta-RB}
\end{figure}

For the qualitative understanding below we use the simplest estimation
(so-called rigid muffin-tin approximation) with the formalism of
Ref.~\cite{MAZINGG}. In this formalism the EPC constant $\lambda$
is proportional to
the Hopfield parameter $\eta $~\cite{MAZINGG}. To calculate $\eta$ one has
to obtain the same complex off-diagonal density matrix at the Fermi level
which enters the expressions for $T_{1}$ (see Section~3). In Fig.~\ref{eta-RB} we show the
behavior of this parameter as a function of $x$ in Mg$_{1-x}$Al$_x$B$_2$
in the rigid band approximation. Changes in $\eta $ are related to
the changes in the electronic $N$, and even neglecting the changes in the
phonon frequencies the general behavior is consistent with the
experimentally observed trend for $T_c$ in this alloy. The $\Gamma$-point
frequencies of the $E_{2g}$ phonon mode obtained in our FLMTO calculation
are 491 cm$^{-1}$ in MgB$_2$ and 956~cm$^{-1}$ in AlB$_2$.
These frequencies are in reasonable agreement with other published data,
as can be seen from Table~\ref{tablephonons}. The biggest disagreement
exists for the most sensitive and important $E_{2g}$ mode where different
calculations produced results ranging from 470 to 665~cm$^{-1}$.
The hardening of the $E_{2g}$ mode with band filling should also
contribute to the suppression of $T_c$ with Al doping.
Still, in spite of the qualitative agreement with experiment and low
anisotropy of in-plane and out-of-plane
contributions to $\eta $, the theoretical $T_{c}$ calculated in the
rigid muffin-tin approximation is too low.

\begin{table}[tbp]
\caption{Calculated phonon frequencies in MgB$_2$ at
$\Gamma$-point (in cm$^{-1}$)}
\begin{center}
\begin{tabular}{|l|l|l|l|l|}
\hline
Authors & E$_{2g}$ & B$_{1g}$ & A$_{2u}$ & E$_{1u}$ \\ 
\hline
Kortus~\cite{us} & 470 & 690 & 390 & 320 \\ 
Yildirim~\cite{nonl} & 486 & 702 & 402 & 328 \\ 
Kunc~\cite{kunc} & 535 & 695 & 400 & 333 \\ 
Present & 491 & 693 & 389 & 326 \\ 
Satta~\cite{Satta} & 665 & 679 & 419 & 328 \\ 
Kong~\cite{ANDERSEN} & 585 & 692 & 401 & 335 \\ 
Bohnen~\cite{Bohnen} & 536 & 692 & 394 & 322\\
\hline
\end{tabular}
\end{center}
\label{tablephonons}
\end{table}

It was suggested~\cite{An} that holes at the top of the B--B bonding
$\sigma$ bands may have strong coupling with the $E_{2g}$ optical B-B
bond-stretching mode. In Figs.~\ref{FermiSurfaces}c and~\ref{FermiSurfaces}d
it is shown how these 2D cylindrical sheets of the Fermi surface breathe
with such distortion (see also Ref.~\cite{Liu}).
The displacements in Figs.~\ref{FermiSurfaces}c and~\ref{FermiSurfaces}d
are (c) $0.07a_B$ and (d) $-0.07a_B$ (positive displacement corresponds to
boron atoms moving towards the centers of hexagons).
The internal cylinder (green in Fig.~\ref{FermiSurfaces}a) falls below $E_F$
and is not seen. It is clearly seen that the bonding $\pi$ band also
experiences strong interaction with the $E_{2g}$ mode.

\begin{figure}[tbp]
\begin{center}
\includegraphics*[scale=0.57,angle=270]{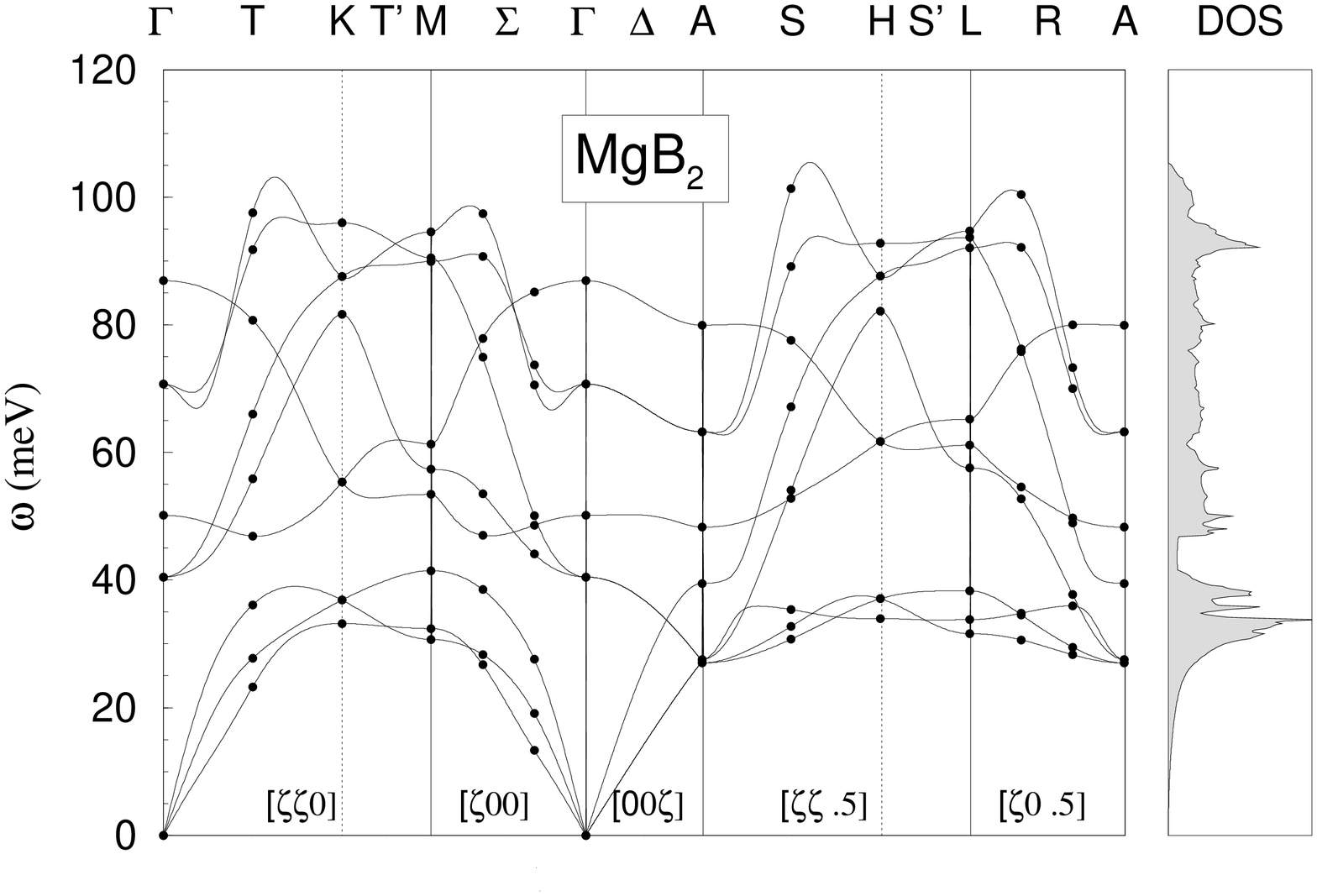}
\includegraphics*[scale=0.57,angle=270]{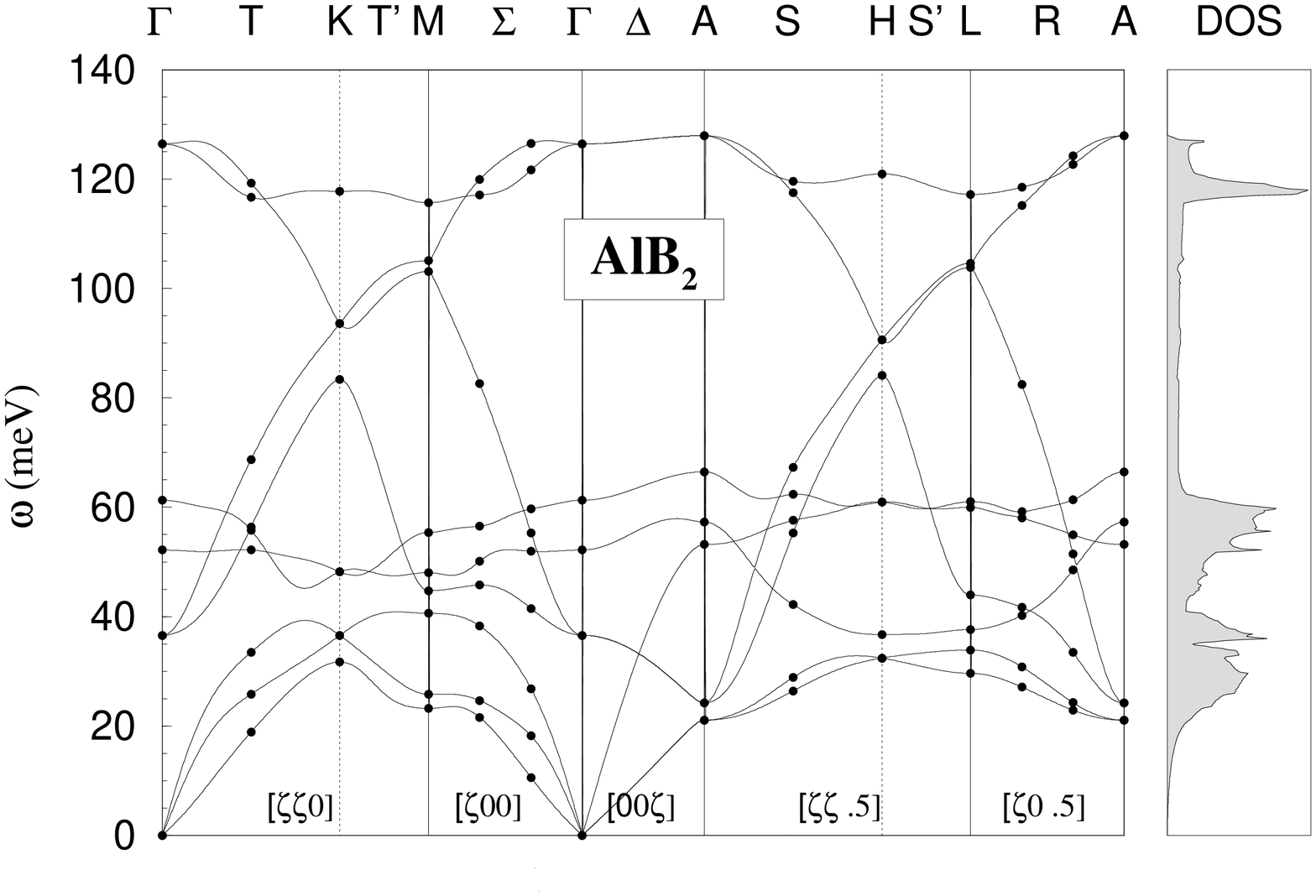}
\end{center}
\caption{Theoretical phonon dispersion curves along the high-symmetry
directions and phonon DOS for MgB$_2$ (top) and AlB$_2$ (bottom).
For details see Ref.~\cite{Bohnen}.}
\label{phonons}
\end{figure}

State-of-the-art calculation of linear response~\cite{ANDERSEN} produced
$\lambda=0.87$, $\omega_{\log }=504$~cm$^{-1}$ and $\mu^{\ast}=0.14$.
The corresponding $T_c$ (from the solution of the Eliashberg equation on the
real axis) is close to 40~K.
The dominance of the $\sigma$--$\sigma$ coupling via the optical
bond-stretching mode was confirmed with the contributions
from $\sigma$ and $\pi $ bands to $\lambda$ amounting to 0.62 and 0.25,
respectively, and the estimated uncertainty of about 6\%. The final
conclusion is that the unusually high $T_c$ is due to the large $\lambda$
caused by the presence of holes in the B--B bonding $\sigma$ bands and
by a relative softness of the $E_{2g}$ mode. It was also concluded that
MgB$_2$ is a clear case of an intermediate-coupling $s$-wave BCS
superconductor.
Other calculations~\cite{Liu,nonl,Bohnen} confirmed
these conclusions.
In Ref.~\cite{nonl} strong non-linearity of
EPC was found and compared with the measured neutron data.
To illustrate the difference between MgB$_2$ and AlB$_2$, in Fig.~\ref{phonons}
we show the phonon spectra from Ref.~\cite{Bohnen} where $\lambda=0.73$
and 0.43 were found for these systems.

In spite of the general agreement between the linear response and frozen
phonon methods,
it is unclear whether these perturbative techniques are
adequate for such strong 
EPC. Unfortunately the detailed analysis of more sophisticated schemes of
pairing~\cite{HIRSCH,Alexandrov} lacks high accuracy of the linear response
method even for this relatively simple material.

Experimental estimation of EPC is not always consistent with the picture
above. If we assume that `bare' LDA $N$ is essentially correct one can use
several experiments to extract the strength of EPC. For instance, comparing
$\gamma$ from the specific heat measurements~\cite{HEAT} with LDA `bare' $N$
according to the relation $\gamma=(1/3)\pi^{2}k_{B}^{2}N(1+\lambda)$
one would obtain $\lambda$ in the range of 0.53--0.62 (the LDA $N$
is~[18-20,~26]
$0.70\pm0.02$ (eV$\cdot$f.u.)$^{-1}$).
With such $\lambda$ an unusually low parameter $\mu^{\ast }=0.03$ should be
used to reproduce the experimentally observed $T_{c}$. In addition, recent
experiments~\cite{GPHONON} revealed no changes in the frequency of the $E_{2g}$
phonon when the system goes through the SC transition.
The analysis of the parameters
$\left( H_{c}\left(0\right) \right) ^{2}/\gamma T_{c}^{2}$
and $\Delta C\left( T_{c}\right)/\gamma T_{c}$
suggests~\cite{HEAT} extremely weak coupling, and the
thermoelectric power data~\cite{TEDS} cannot be fit with the LDA $N$.
These controversies leave
a question whether the electron pairing in MgB$_{2}$ is phonon-mediated
relatively open. We should stress that there is a large disagreement
between different measurements. For example, the  experimental
data~\cite{otherheat} for the specific heat $\gamma$ range
from 1.1 to 5.5 mJ mol$^{-1}$ K$^{-2}$. 
Analyzing $N$ extracted from NMR data (see Section~3) one can find that the
LDA partial boron $N$ should be lowered by 10--15\% to comply with the $T_1$
measurements. With such correction for the total $N$ the corrected
$\lambda$ can easily be equal to~0.8. Thus, a lower $N$ is required to
make a satisfactory interpretation of these two experiments. At present
the reasons for lowering of $N$ are unclear. In any case the application
of many-body techniques (GW or others) to this material should be valuable.

\section{Dielectric function and reflectivity}

Since optical properties reflect the electronic structure of a system, it is
of interest to interpret the main features of the optical spectra in terms
of the electronic transitions in the studied materials.
Here we are dealing with the
optical properties as expressed by the frequency-dependent dielectric
function associated with direct interband transitions.

The imaginary part $\varepsilon_{2}(\omega)$ in the random phase
approximation (RPA) with the LMTO basis set is~\cite{Alouani}: 
\[
\varepsilon _{2}^{j}(\omega )=\frac{e^{2}}{m^{2}\omega ^{2}\pi }%
\sum_{\lambda ,\lambda ^{\prime }}\int_{BZ}d{\bf k}\;|
\langle{\bf k}\lambda|-i\hbar\partial_{j}|{\bf k}\lambda^{\prime}\rangle|^{2}
\;f_{\lambda }^{{\bf k}}(1-f_{\lambda ^{\prime }}^{{\bf k}})\;\delta (\epsilon
^{\bf k}_{\lambda}-\epsilon^{\bf k}_{\lambda^{\prime}}-\hbar\omega), \label{e2} 
\]
where $\epsilon^{\bf k}_{\lambda}$ and $\epsilon^{\bf k}_{\lambda^{\prime}}$
are the eigenvalues for bands $\lambda$ and $\lambda^{\prime}$,
respectively, and $f_{\lambda }^{{\bf k}}$ is
the occupation number at zero temperature. For AlB$_{2}$ structure, there
are two independent components for ${\bf E}\perp$c and ${\bf E}\parallel$c.

The complex index of refraction is given by
\[
\tilde{n}=n+ik=\varepsilon ^{1/2}=(\varepsilon_{1}+i\varepsilon _{2})^{1/2}\,,
\label{eq:refrac} 
\]
and the normal incidence reflectivity if defined by the Fresnel equation 
\[
R=\frac{(n-1)^{2}+k^{2}}{(n+1)^{2}+k^{2}}\;.\label{eq:reflec} 
\]

We calculated the optical response for both polarizations. The tetrahedron
method was used for the Brillouin zone integrations using 512
irreducible ${\bf k}$ points. The real part $\varepsilon _{1}\left( \omega
\right) $ was obtained from $\varepsilon _{2}\left( \omega \right) $ by the
Kramers-Kronig relation. The $\varepsilon \left( \omega \right) $ functions
were calculated up to 25 eV, which is well above the main peak of $%
\varepsilon _{2}\left( \omega \right) $. Also, both in MgB$_{2}$ and AlB$%
_{2} $ there are partially filled bands, and therefore, one should take into
account the intraband, or Drude, term. The plasma frequency in the Drude
term is calculated from the average square of the Fermi velocity
(see Section~2), and the scattering rate was obtained from the
experimental data on resistivity in the normal state. This procedure
is quite standard and was 
used for different metals including high-$T_{c}$ superconductors~\cite
{Our1,Our2}. Since we have no accurate ellipsometric measurements of the
dielectric function for these materials, our results should be considered as
a prediction.

\begin{figure}[ptb]
\begin{center}
\includegraphics*[scale=0.6]{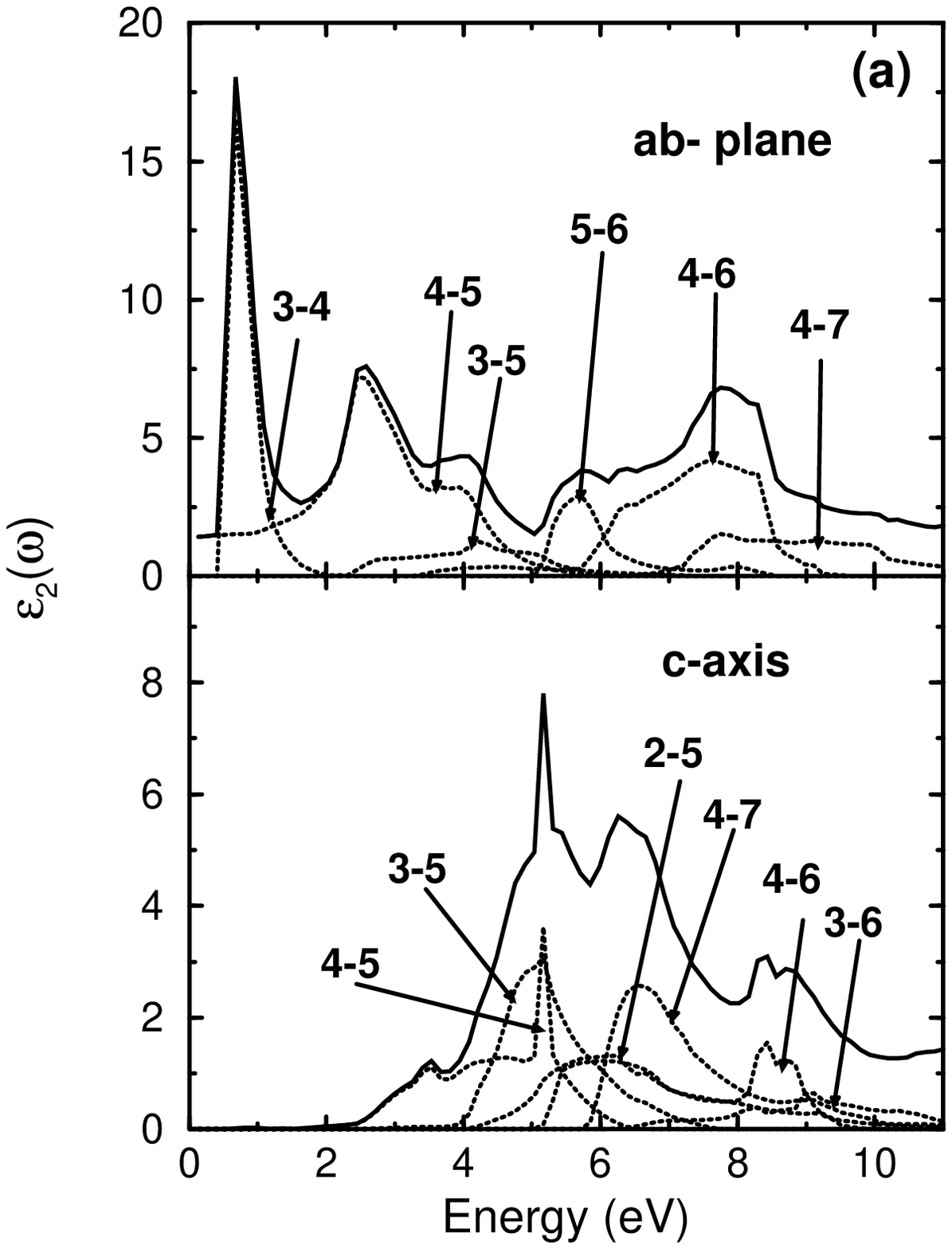}
\includegraphics*[scale=0.6]{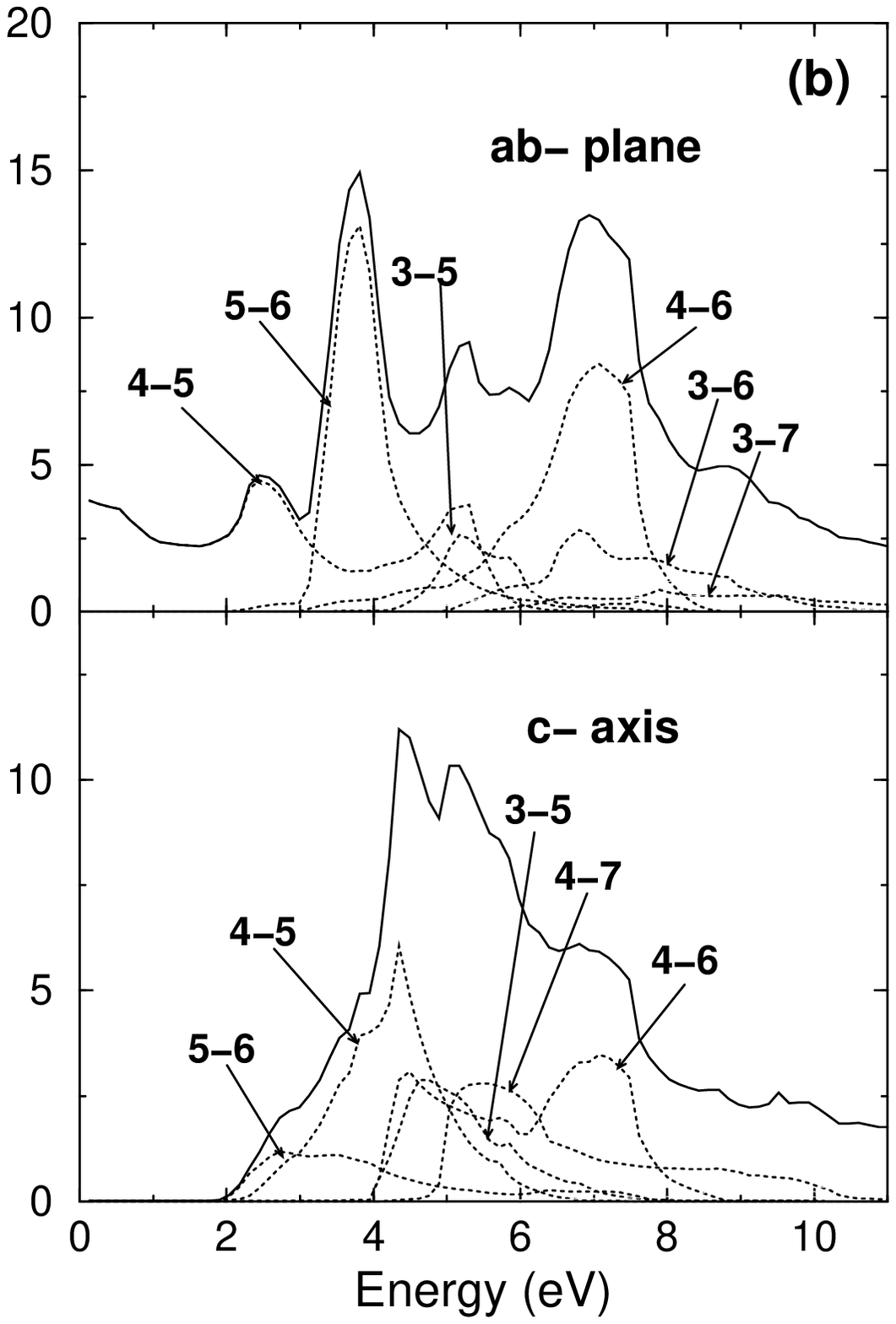}
\end{center}
\caption{Calculated $\varepsilon(\omega)$ for (a) MgB$_2$ and (b) AlB$_2$
with its band-to-band decomposition.}
\label{diel}
\end{figure}

Fig.~\ref{diel} shows the interband contributions to
$\varepsilon_{2}(\omega)$ for $\omega<11$~eV for both materials and
both polarizations.
Also, we show the partial band-to-band contributions to the imaginary
part of the dielectric tensor for all significant transitions.

Both ${\bf E}\perp $c and ${\bf E}\parallel $c components are very
anisotropic and quite different for both materials. In MgB$_{2}$, the
function $\varepsilon _{2}\left( \omega \right) $ exhibits a structure with
two broad peaks in ${\bf E}\perp $c. The maximum at 2.5 eV corresponds to
the transitions between the 4-th and 5-th bands mostly around the
$\Gamma$--M direction at the Brillouin zone (see Fig.~\ref{mgb2bnds}).
The second maximum at about 8~eV is related to the 4$\leftrightarrow$6
transitions related to the H--A direction where these bands contain
extended nearly parallel parts. For the polarization ${\bf E}\parallel $c
there is only one broad hump with several fine features which can be
easily identified using the band-to-band transition analysis
(Fig.~\ref{diel}a). The sharp peak at 5~eV is related to the
4$\leftrightarrow$5 transitions from narrow region around H--A direction
where these bands are also nearly parallel.

In AlB$_{2}$ (Fig.~\ref{diel}b) the two main peaks for the $ab$
polarization
are related to the 5$\leftrightarrow$6 (3.8~eV peak) and 4$\leftrightarrow$6
(7~eV peak) transitions, respectively. For
the $c$ polarization there is (like in MgB$_{2}$) a single broad hump
with the maximum at about 4.2 eV and with several noticeable features that
are easily identifiable.

There are two main differences between the MgB$_{2}$ and AlB$_{2}$ interband 
$\varepsilon _{2}\left( \omega \right) $ spectra. First, we notice that
there is a low-energy peak (0.7 eV) in MgB$_{2}$ which is absent in
AlB$_{2}$. This peak is related to the transitions between bands~3 and 4
in the H-A direction. The difference of energies for these two bands is very
low in this direction. In MgB$_{2}$ these bands cross the Fermi level while
in AlB$_{2}$ they are both below the Fermi energy, i.e., this peak does not
appear.

Second, the absolute values of $\varepsilon
_{2}\left( \omega \right) $ are higher in AlB$_{2}$ compared to MgB$_{2}$ in
all the considered energy interval and for both polarizations. The analysis
of the joint density of states (which is equal to $\varepsilon_{2}(\omega)$
when all matrix elements are set to unity) shows that it is
related to larger matrix elements in AlB$_{2}$. The reason
for such behavior is unclear, and it is necessary to perform a detailed
analysis of the wave functions for the bands surrounding the Fermi energy.

\begin{figure}[tbp]
\begin{center}
\includegraphics*[scale=0.84]{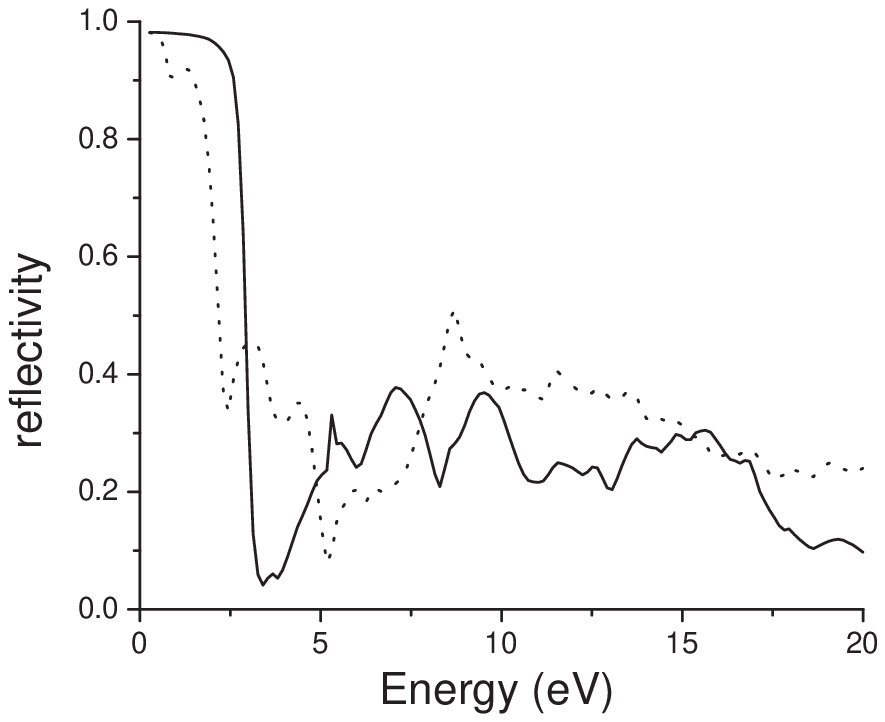}
\includegraphics*[scale=0.84]{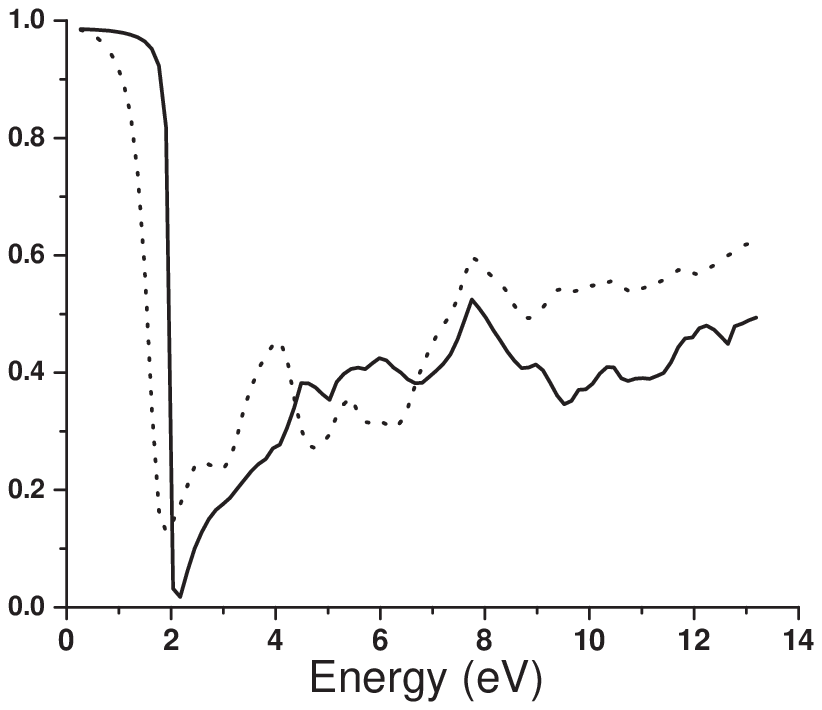}
\end{center}
\caption{Calculated reflectivity for MgB$_{2}$ (left) and AlB$_{2}$ (right).
Solid lines, $c$ polarization; dotted lines, $ab$ polarization.}
\label{refl}
\end{figure}

The reflectivity spectra for MgB$_{2}$ and AlB$_{2}$ are shown in
Fig.~\ref{refl}. These spectra are anisotropic. For MgB$_{2}$ the
anisotropy of the plasma frequency is not very high ($\omega _{p}^{ab}$= 6.5
eV, $\omega _{p}^{c}$=7.02 eV). However, the anisotropy of the reflectivity
spectra is high even in the low-energy region due to interband transitions. For
example, for the $ab$ polarization there is a feature at
$\sim1.3$~eV corresponding to the peak of the interband part of $\varepsilon_2(\omega)$
at $\sim0.7$~eV (Fig.~\ref{refl}). We believe that
this structure can be observed in the reflectivity measurements. AlB$_{2}$
does not have this low-energy peak, and therefore the behavior of the
reflectivity at the energies below 2~eV is rather smooth.

The studies of the general dielectric function $\varepsilon({\bf k},\omega)$
and analysis of collective charge excitations for MgB$_{2}$ appeared
recently in Ref.~\cite{EPS}.

\section{Conclusion}

We showed that the electronic structure of Mg$_{1-x}$Al$_{x}$B$_{2}$ has
both similarity with and notable differences to GIC's. Its peculiar
features include holes at the top of the $\sigma $ bands, coexistence of
metallic and covalent bonding, and smooth disappearance of the 2D character in
the electronic structure with Al doping. We emphasize how new NMR and
optical experiments may provide a direct test of validity of these
theoretical predictions. The presented results provide a basis both for
further studies of normal and SC states of MgB$_{2}$ and for the search
of new SC compounds. Such studies are likely to be very complicated due to the
fact that MgB$_{2}$ seems to be a unique superconducting compound among the
family of structurally similar materials with no or very low-$T_c$ SC. It is
unlikely that an analog to the high-$T_c$ family of materials can be
found in the case of MgB$_2$. Nevertheless, the relative simplicity of
this compound makes the development of new computational schemes with
realistic electron-phonon coupling very attractive.

The authors acknowledge useful discussions with F.~Borsa, S.~Bud'ko,
P.~Canfield, I.~Mazin and N.~Zein. This work was carried out at the Ames
Laboratory, which is operated for the U.S. Department of Energy by Iowa
State University under Contract No. W-7405-82. This work was supported by
the Director for Energy Research, Office of Basic Energy Sciences of the
U.S. Department of Energy.

\end{document}